\documentclass[12pt]{article}

\usepackage[dvips]{graphics,epsfig}
\usepackage[latin1]{inputenc}
\usepackage{setspace}
\usepackage{amsfonts}
\usepackage{cite}
\usepackage{textcomp}
\usepackage{amsmath}
\usepackage{graphicx}
\usepackage{float}
\usepackage{braket}
\usepackage{subfig}
\usepackage{dsfont}
\graphicspath{ {./plots/} }
\def\unit{\mathds{1}}

\newcommand{\realp}{\operatorname{Re}}

\usepackage[unicode=true,bookmarks=false,breaklinks=false,pdfborder={0 0 1},colorlinks=true]
 {hyperref}
\hypersetup{
 citecolor=blue,linkcolor=blue,urlcolor=blue}

\def\laq{\raise 0.4 ex \hbox{$<$}\kern -0.8 em\lower 0.62 ex\hbox{$\sim$}}
\def\gaq{\raise 0.4 ex \hbox{$>$}\kern -0.7 em\lower 0.62 ex\hbox{$\sim$}}

\def\beq{\begin{equation}}
\def\eeq{\end{equation}}
\def\beqa{\begin{eqnarray}} 
\def\eeqa{\end{eqnarray}}
\begin{document}

\pagestyle{plain}

\begin{flushright}
{\bf DRAFT VERSION}\\
May 23, 2022
\end{flushright}
\vspace{0.2cm}

\begin{center}

{\Large\bf Entanglement dynamics: Generalized master equation \\ for uniformly accelerated two-level systems}

\vspace*{0.5cm}

M. S. Soares$^{*}$, N. F. Svaiter$^{\dagger}$\\
{Centro Brasileiro de Pesquisas F\'{\i}sicas\\
Rua Xavier Sigaud, 150 - Urca, Rio de Janeiro - RJ, 22290-180, Brazil}\\

\vspace*{0.5cm}

G. Menezes$^{\wr}$\\
{Universidade Federal Rural do Rio de Janeiro,\\
 Rodovia BR 465, Km 07,\\
Serop\'edica-RJ, - Brazil}\\
\end{center}

\begin{abstract}
We propose a new form for the quantum master equation in the theory of open quantum systems. This new formalism allows one to describe the dynamics of two-level systems moving along different hyperbolic trajectories with distinct proper times. In the Born-Markov approximation, we consider a quantum massless scalar field coupled with two-level systems. Starting from a separable state we show the emergence of entanglement harvesting. For different proper accelerations we verify also the entanglement sudden death.
\end{abstract}

\section{Introduction}

In a generic curved spacetime, a natural decomposition of modes in terms of negative and positive frequency parts is not generally available. The non-uniqueness of the vacuum state is an important consequence of this circumstance~\cite{Bire82,Parker:2009uva}. Even for observers in Minkowski spacetime, there are non-trivial situations involving the choice of a vacuum state~\cite{Davies:1974th,Unruh:1976db,unruh1984}. For instance, a uniformly accelerated observer in flat spacetime perceives the Minkowski vacuum as a thermal bath. This is known as the Unruh-Davies effect and has a close connection to the Hawking effect \cite{hawking75, yu08}. The discussion of quantum fields in non-inertial frames and related subjects has been a topic under vigorous investigation over the years \cite{fulling1973,Candelas:1976jv,svaiter1992,Ford:1994zz,Shih-Yuin:06,Crispino:2007eb,Lenz:2010vn,Arias:2011yg,Menezes:2015iva,Ben-Benjamin:2019opz,Rodriguez-Camargo:2016fbq, Arias:2015moa,Fewster:2016ewy,Menezes:2016quu, Menezes:2017rby, Picanco:2020api, PhysRevD.101.025009,Tjoa:2022oxv,Hu:2022nxc,kaplanek2020hot,arrechea2021inversion,carballo2019unruh,juarez2014onset, PhysRevD.95.025020}.

An apparatus device in such studies is the Unruh-deWitt particle detector \cite{dewitt1975,Hawking:1979ig}. It is defined as an idealized two-level system  coupled with a scalar field through a monopole interaction. Other descriptions for particle detectors are also possible as, for example, the one given by the Glauber theory~\cite{glauber1963A}. A recent discussion on particle-detector models can be found in Ref.~\cite{soares21}. In this work, the authors have highlighted the importance of the Glauber model in the clarification on the different interpretations of the measurements performed by an accelerated detector. See also Refs.  \cite{Martin-Martinez:2014qda,Martin-Martinez:2015psa,Hummer:2015xaa,Tjoa:2021bcn,deRamon:2021nry}. 

An essential tool in quantum computation and quantum information theory is quantum entanglement \cite{nielsen2011}. The literature has examined several sources of entangled quantum systems, recurrently found in solid-state physics and atoms in cavity electrodynamics. In fact, a number of proposals to generate entangled states in systems of two-level systems coupled to bosonic fields were established \cite{Plenio:1998wq,cite-key,doi:10.1080/09500340308234584,Tana_2004,Amico:2007ag}.

In recent years, quantum entanglement has been analysed through the lens of the relativistic quantum information theory, which studies the behaviour of atoms interacting with relativistic quantum fields~\cite{Peres:2002ip,Gingrich:2002ota,Gingrich:2002otaa,Shi:2004yt,Czachor,Jordan:2006kt,Adesso:2007wi,Datta,Peres:2002wx,Lin:2010zzb,Bruschi:2010mc,Ostapchuk:2011ud,Hu:2012jr,CQG12,Lin:2015aua}. A well known phenomenon  in this scenario is the so-called entanglement degradation, when correlated states become uncorrelated by the interaction with a quantum field~\cite{Martin-Martinez:2010yva,yu2016pra,She:2019hjv,He:2020xhz,Bhattacharya:2021zgd}. Another effect is the entanglement harvesting. Atoms initially prepared in a separable state can extract entanglement from the quantum vacuum~\cite{martinmartinez-PhysRevD.92.064042,Sachs:2017exo,Pozas-Kerstjens:2017xjr,MM-PhysRevD.94.064074,Liu:2020jaj,Tjoa:2020riy,Perche:2021clp}. Vacuum fluctuations are able to act as a source of entanglement for atoms coupled with quantum fields. This can be realized, for instance, when atoms are uniformly accelerated, as the Minkowski vacuum state can be conceived as a multi-particle state with Rindler excitations~\cite{unruh1984}. In any case, it is important to bear in mind that entanglement harvesting protocol can be considered as legitimate only when the detectors are not able to communicate \cite{martinmartinez_prd21}.

Relativistic quantum entanglement has been investigated in different setups~\cite{Fuentes-Schuller:2004iaz,mann2008quantum, Alsing_2012, Salton_2015, PhysRevD.104.025001, PhysRevD.92.064042, PhysRevA.100.062126, PhysRevD.102.065013, cong2020effects, liu2021does,Menezes:2015uaa, Menezes:2015veo,Menezes:2017oeb}. Recently Benatti and Floreanini discussed entanglement dynamics of two non-inertial atoms weakly interacting with a quantum scalar field with the same proper acceleration \cite{benatti04_pra}.  See also Ref. \cite{benatti2010,louko_prd_2020}. For the case of atoms also with the same proper acceleration but interacting with the electromagnetic field see Ref. \cite{hu2015}. In this work we study a pair of two-level systems travelling along two different worldlines. We generalize the construction of the \textit{master equation} widely discussed in the literature \cite{breuer2002theory, tutorial, cohen1986quantum}. Our formalism allows one to analyse the implications of different accelerations for entanglement degradation and entanglement harvesting. In order to quantify the content of entanglement between the two-level system, we calculate the concurrence introduced by Wootters \cite{wootters98}.

The organization of this paper is as follows. In Sec. \ref{sec:thegme} we present the derivation of the master equation for two-level systems travelling in different world lines. In  Sec. \ref{sec:thegme_coupleofatoms} we apply this equation for two two-level systems interacting with a scalar field. In Sec. \ref{sec:ent_dynamics} we discuss entanglement dynamics. Conclusions are given in Sec. \ref{sec:conclusions}. We use units such that $\hbar = c = k_{B} = 1$.

\section{Generalized Master Equation}\label{sec:thegme}

In this section we derive a generalized form of the master equation \cite{tutorial}. The term \textit{generalized} is been used since we are working with pointlike two-level systems travelling along different world lines with different proper times. The Hamiltonian of the system that we are interested in can be written as
	\begin{equation}\label{eqq:hsys}
	\mathcal{H} = \mathcal{H}_{0} + \mathcal{H}_{\text{int}} = \mathcal{H}_{A} + \mathcal{H}_{F} + \mathcal{H}_{\text{int}}.
	\end{equation}
In the above equation, $\mathcal{H}_A$ describes the Hamiltonian of the two-level system, $\mathcal{H}_F$ is the scalar field Hamiltonian and $\mathcal{H}_{int}$ describes the interaction Hamiltonian. Usually, the interaction Hamiltonian $( \mathcal{H}_{\text{int}})$	has the form
	\begin{equation}
	 \mathcal{H}_{\text{int}} = \sum_{i}\sum_{\alpha} A^{(\alpha)}_{i} \otimes B^{(\alpha)}_{i},
	\end{equation}
where $A_i^{(\alpha)}$ are operators which act in the space of the states of two-level systems, while operators $B_i^{(\alpha)}$ correspond to space of the field's states.  The $\alpha$ index denotes the number of two-level systems.

In order to describe the evolution of the subsystem $A$ one must choose an appropriate parameter. In our exposition, one can expect to employ the proper time of the two-level systems, as the coupling with the quantum field is effective only on their trajectory. Nevertheless here there is an ambiguity -- the two-level systems under studied can move along different trajectories parametrized by different proper times. In this case, one can imagine the existence of an auxiliary ``time parameter"~$\eta$ (in the sense that it is an arbitrary parametrization of a curve describing the world line of an ``atom" or a ``particle detector") such that one can make the following reasonable assumption: The proper times of the two-level systems are functions of this quantity, $\tau_{\alpha} = \tau_{\alpha}(\eta)$. What this parameter would be will depend on the context under analysis. The reason that we initially left $\eta$ unspecified is that we wish to show that the formalism put forward in this paper can cover many different situations that obey the aforementioned general condition. Accordingly, an appropriate and unambiguous definition for $\eta$ depends on the situation in which one is interested. Of course, in a curved spacetime the notion of a coordinate time is more subtle and one must be careful in their choice -- simply stated, guidelines in this situation should be considered on a case by case basis. Nevertheless, as we will discuss in the next section, we are interested in the situation in which the two-level systems are in uniformly accelerated world lines. In this case $\eta$ can be the associated Rindler time. In fact, since two different hyperbole in the same Rindler wedge can be cut by a single line of constant $\eta$, such proper times can be simply related, and this feature will be exploited in due course.

For the case of two-level systems the interaction Hamiltonian in the interaction picture can usually be written as
	\begin{align}
	 \mathcal{H}_{\text{int}}(\eta) &= \sum_{\omega}\sum_{i}\sum_{\alpha} e^{-i\omega \tau_{\alpha}(\eta)}   A^{(\alpha)}_{i}(\omega)\otimes B^{(\alpha)}_{i}\left[x(\tau_{\alpha}(\eta))\right]\frac{d\tau_{\alpha}(\eta)}{d\eta},\label{eqq:inthamilton}
	\end{align}
where $\omega$ is the energy gap between the two-level states. The density matrix of the whole system $(\rho_{AB})$ in the interaction picture $(\tilde{\rho}_{AB})$ is described by the Liouville-von Neumann equation
	\begin{equation}\label{eqq:vonneuman}
	\frac{d}{d\eta} \tilde{\rho}_{AB}(\eta) = - i \left[\mathcal{H}_{\text{int}}(\eta),\tilde{\rho}_{AB}(\eta) \right] .
	\end{equation}
Although this equation gives the dynamics of the whole system, we are interested only in the behavior of the subsystem $A$. Our aim is to construct an equation for the reduced density matrix 
$\rho_{A}(\eta)$ of the two-level systems which preservers all the properties of the density matrices.

By making use of perturbation theory with respect to a suitable coupling constant, which is present in the interaction Hamiltonian, and considering the weak-coupling approximation, $i.e.$, neglecting higher-order correlations,  we find that 
	\begin{align}
	\Delta \tilde{\rho}_{AB}(\eta)& = -i \int_{\eta}^{\eta + \Delta \eta}d\eta_1 \left[\mathcal{H}_{\text{int}}(\eta_1),\tilde{\rho}_{AB}(\eta_1) \right]\nonumber\\ &- \int_{\eta}^{\eta + \Delta \eta}d\eta_1 \int_{\eta}^{\eta_1} d\eta_2 \left[\mathcal{H}_{\text{int}}(\eta_{1}),\left[\mathcal{H}_{\text{int}}(\eta_{2}),\tilde{\rho}_{AB}(\eta) \right] \right]. \label{eqq:weakapprox}
	\end{align}		
As we wish to study the evolution of the reduced density matrix  $\rho_A$ of the subsystem ${\cal A}$, we trace over the field states, $\tilde{\rho}_{A}(\eta) = \textrm{Tr}_{B}\left[ \tilde{\rho}_{AB}(\eta) \right]$, and obtain that
	\begin{align}
	\Delta \tilde{\rho}_{A}(\eta) &= -i \int_{\eta}^{\eta + \Delta \eta}d\eta_1~ \textrm{Tr}_{B}\left[\mathcal{H}_{\text{int}}(\eta_1),\tilde{\rho}_{AB}(\eta_1) \right]\nonumber\\ &- \int_{\eta}^{\eta + \Delta \eta}d\eta_1 \int_{\eta}^{\eta_1} d\eta_2  \textrm{Tr}_{B}\left[\mathcal{H}_{\text{int}}(\eta_{1}),\left[\mathcal{H}_{\text{int}}(\eta_{2}),\tilde{\rho}_{AB}(\eta) \right] \right]. \label{eqq:weakapprox_partial}
	\end{align}		
Before proceeding, we must resort to some additional approximations. We would like to stress that the derivation presented below depends strongly on these assumptions. This procedure is justified as such approximations can be implemented in many situations well described in the literature. 

We assume that there are two distinct time scales. One is specified by $\tau_{\cal B}$, the characteristic time associated with the internal correlations in the system ${\cal B}$. The other time scale is $\tau_{\cal A}$ which describes the time evolution of the density matrix in the presence of the interaction between the two systems. We assume that
	\begin{equation}
	\tau_{\cal B} \ll \Delta \eta \ll \tau_{\cal A}.
	\end{equation}

This means that the time scale relaxation of the system ${\cal B}$ is smaller then the time scale relaxation of the two-level systems. Therefore, when the correlation between the two systems starts, the field has enough time to reach the equilibrium. This implies that we are within the regime of the \textit{Born approximation}, which enables one to neglect initial correlations induced by the interactions between the subsystems:
	\begin{equation}\label{eqq:born_approx}
	\tilde{\rho}_{AB}(\eta) \approx \tilde{\rho}_{A}(\eta) \otimes \tilde{\rho}_{B},
	\end{equation}	 	
where $\tilde{\rho}_{B}$ is the reduced density operator associated with subsystem ${\cal B}$. Furthermore, such approximations also allow us to conclude that the operator $\tilde{\rho}_{B}$ can be approximately regarded as being stationary:
	\begin{equation}
	\tilde{\rho}_{B}(\eta) \approx \sigma_{B} \rightarrow \left[\sigma_{B}, \mathcal{H}_{B}\right] = 0.
	\end{equation}	 
As mentioned above, the subsystem ${\cal B}$ is associated with the quantum field to which the two-level systems couple. Therefore, one expect that
	\begin{equation}
	\textrm{Tr}_{B}\left\lbrace{B^{(\alpha)}_{i}\left[x(\tau_{\alpha}(\eta))\right]\sigma_{B}}\right\rbrace = 0.
	\end{equation}		
So we can rewrite Eq. \eqref{eqq:weakapprox_partial} as
	\begin{align}
	\Delta \tilde{\rho}_{A}(\eta) = &- \int_{\eta}^{\eta + \Delta \eta}d\eta_1 \int_{\eta}^{\eta_1} d\eta_2\nonumber\\
	&\times \textrm{Tr}_{B}\left[\mathcal{H}_{\text{int}}(\eta_{1}),\left[\mathcal{H}_{\text{int}}(\eta_{2}),\tilde{\rho}_{AB}(\eta) \right] \right], \label{eqq:weakapprox_partial_appro}
	\end{align}		

Expanding the commutators and employing the aforementioned discussion of time intervals, we find that 
	\begin{align}
		\frac{\Delta \tilde{\rho}_{A}(\eta)}{\Delta \eta} &= \frac{1}{\Delta \eta} \int_{\eta}^{\eta + \Delta \eta}d\eta_1 \int_{\eta}^{\eta_1} d\eta_2 \textrm{Tr}_{B}\left[ \mathcal{H}_{\text{int}}(\eta_{2})\tilde{\rho}_{A}(\eta)\otimes \sigma_{B} \mathcal{H}_{\text{int}}(\eta_{1})\right. \nonumber\\
		&- \left.\mathcal{H}_{\text{int}}(\eta_{1})\mathcal{H}_{\text{int}}(\eta_{2})\tilde{\rho}_{A}(\eta)\otimes \sigma_{B}   \right] + \textrm{h.c.} \label{eqq:weakapprox_partial_2}
	\end{align}
Using Eq.~\eqref{eqq:inthamilton}:
	\begin{align}
	\frac{\Delta \tilde{\rho}_{A}(\eta)}{\Delta \eta} &= \frac{1}{\Delta \eta}\sum_{\omega, \omega^{'}}\sum_{i,j}\sum_{\alpha, \beta} \int_{\eta}^{\eta + \Delta \eta}d\eta_1 \int_{\eta}^{\eta_1} d\eta_2~e^{- i \omega \tau_{\beta}(\eta_2)}e^{i \omega^{\prime} \tau_{\alpha}(\eta_{1})}\nonumber\\
	&\times \left(\left. A^{(\beta)}_{j}(\omega)\tilde{\rho}_{A}(\eta) A^{\dagger (\alpha)}_{i}(\omega^{\prime})\textrm{Tr}_{B}\left[ B^{(\beta)}_{j}(\eta_{2})\sigma_{B}B^{\dagger(\alpha)}_{i}(\eta_{1})\right]\right. \right.\nonumber \\ 
	 &\left. -  A^{\dagger (\alpha)}_{i}(\omega^{\prime}) A^{(\beta)}_{j}(\omega)\tilde{\rho}_{A}(\eta)\textrm{Tr}_{B}\left[ B^{\dagger(\alpha)}_{i}(\eta_{1}) B^{(\beta)}_{j}(\eta_{2})\sigma_{B}\right]\right)\frac{d \tau_{\beta}(\eta_1)}{d\eta_{1}}\frac{d \tau_{\alpha}(\eta_2)}{d\eta_{2}} + \textrm{h.c.},  
	\end{align}
where for brevity we wrote $B^{(\alpha)}_{i}\left[x(\tau_{\alpha}(\eta))\right] = B^{(\alpha)}_{i}(\eta)$. We define the field correlation function as
	\begin{equation}\label{eqq:def_whightman_trace}
	\textrm{Tr}_{B}\left[B^{\dagger(\alpha)}_{i}(\eta_{1}) B^{(\beta)}_{j}(\eta_{2})\sigma_{B}\right] = G^{(\alpha \beta)}_{ij}(\tau_{\alpha}(\eta_1), \tau_{\beta}(\eta_2)).
	\end{equation}
One can write that 
	\begin{align}
	G^{(\alpha \beta)}_{ij}(\tau_{\alpha}(\eta_1), \tau_{\beta}(\eta_2)) = \textrm{Tr}_{B}\left[e^{ - i \mathcal{H}_{F} \tau_{\alpha}(\eta_1) }B^{\dagger(\alpha)}_{i}(0)e^{ i \mathcal{H}_{F} \tau_{\alpha}(\eta_1) }e^{ - i \mathcal{H}_{F} \tau_{\beta}(\eta_2) } B^{(\beta)}_{j}(0)e^{  i \mathcal{H}_{F} \tau_{\beta}(\eta_2) }\sigma_{B}\right],
	\end{align}	
and by making use that $\sigma_{B}$ does not change in time and the cyclic property of the trace we have
	\begin{align}
	G^{(\alpha \beta)}_{ij}(\tau_{\alpha}(\eta_1), \tau_{\beta}(\eta_2)) = \textrm{Tr}_{B}\left[B^{\dagger(\alpha)}_{i}\left(x\left[\tau_\alpha(\eta_1) - \tau_\beta(\eta_2)\right]\right) B^{(\beta)}_{j}(0)\sigma_{B}\right].
	\label{eqq:diff_prop_time}
	\end{align}
As one can observe by Eq. \eqref{eqq:diff_prop_time} the function $G^{(\alpha \beta)}_{ij}(\tau_{\alpha}(\eta_1), \tau_{\beta}(\eta_2))$ only depends in the difference  $\tau_{\alpha}(\eta_1) - \tau_{\beta}(\eta_2)$. In many cases that we are interested, this function is the positive frequency Wightman function. Therefore we get in the argument the difference of the proper times of the two-level systems. As will be discussed in the next section, in the present work we are particularly interested in uniformly accelerated two-level systems, and in this case we will see that this function only depends on the difference of the parameters $\eta_1 - \eta_2$, which here will be taken to be the proper time of one of the two-level systems.  Therefore we write that
	\begin{equation}
		G^{(\alpha \beta)}_{ij}(\tau_{\beta}(\eta_1), \tau_{\alpha}(\eta_2)) = 
		G^{(\alpha \beta)}_{ij}(\eta_1 - \eta_2).
	\end{equation}
Using the cyclic property of the trace, and after a simple algebraic manipulation, we find that
	\begin{align}
	\frac{\Delta \tilde{\rho}_{A}(\eta)}{\Delta \eta} =&\frac{1}{\Delta \eta} \sum_{\omega, \omega^{'}}\sum_{i,j}\sum_{\alpha, \beta} \int_{\eta}^{\eta + \Delta \eta}d\eta_1 \int_{\eta}^{\eta_1} d\eta_2~e^{i(\omega^{\prime} - \omega)\tau_{\alpha}(\eta_{1})} e^{i \omega\left(\tau_{\alpha}(\eta_{1}) - \tau_{\beta}(\eta_{2})\right)}\nonumber\\
	 &\times G^{(\alpha \beta)}_{ij}(\eta_1 - \eta_2)\frac{d \tau_{\beta}(\eta_1)}{d\eta_{1}}\frac{d \tau_{\alpha}(\eta_2)}{d\eta_{2}} \left( A^{(\beta)}_{j}(\omega)\tilde{\rho}_{A}(\eta) A^{\dagger (\alpha)}_{i}(\omega^{\prime})\right.\nonumber\\
	 & -\left.  A^{\dagger (\alpha)}_{i}(\omega^{\prime}) A^{(\beta)}_{j}(\omega)\tilde{\rho}_{A}(\eta)\right) + h.c.\label{eqq:gme_nochange}  
	\end{align}
Up to now we have not specified the behaviour of the functions $\tau_{\alpha}(\eta)$. For simplicity, let us assume a linear relation of the form
	\begin{equation}\label{eqq:a-deff}
	\tau_{\alpha}(\eta_{i}) = a_{\alpha}\eta_{i},
	\end{equation}
for some constant $a_{\alpha}$. This relation is obeyed in several interesting physical situations, as, for example, for some classes of observers in Rindler and Schwarzschild spacetimes -- in the former case, $a_{\alpha}$ can be associated with the proper acceleration of the two-level systems. 

Defining  $s = \eta_{1} - \eta_{2}$\footnote{Note that $\alpha$ and $\beta$ indices define two distinct worldlines of the atoms and $\eta_1$ and $\eta_2$ define two hypersurfaces of simultaneity. Therefore $\alpha = \beta$ does not implies $s = \eta_1 - \eta_2 =0$.}, and performing the following change of coordinates (for a futher analysis about this change of coordinates see Ref. \cite{tutorial})
	\begin{align}
	 \int_{\eta}^{\eta + \Delta \eta}d\eta_1 \int_{\eta}^{\eta_1} d\eta_2 \rightarrow \int_{0}^{\Delta \eta} ds \int_{\eta +s}^{\eta +\Delta \eta}d\eta_1,
	\end{align}
we obtain
	\begin{align}
	\frac{\Delta \tilde{\rho}_{A}(\eta)}{\Delta \eta} &=\sum_{\omega, \omega^{'}}\sum_{i,j}\sum_{\alpha, \beta}J(\omega^{\prime} a_{\alpha} - \omega a_{\beta}) W_{ij}^{(\alpha \beta)}(a_{\beta} \omega) 
	 \left( A^{(\beta)}_{j}(\omega)\tilde{\rho}_{A}(\eta) A^{\dagger (\alpha)}_{i}(\omega^{\prime}) \right.\nonumber\\
	 &-\left.  A^{\dagger (\alpha)}_{i}(\omega^{\prime}) A^{(\beta)}_{j}(\omega)\tilde{\rho}_{A}(\eta)\right)
	 + \textrm{h.c.},\label{eqq:gme_nochange_3}  
	\end{align}
where we have defined the following functions
	\begin{equation}\label{eqq:J_function}
	J(\omega^{\prime} a_{\alpha} - \omega a_{\beta}) = \int_{\eta}^{\eta + \Delta \eta}d\eta_{1}~\frac{e^{i(\omega^{\prime} a_{\alpha} - \omega a_{\beta})\eta_{1}}}{\Delta \eta},
	\end{equation}
and
	\begin{equation}\label{eqq:w_function_general}
		W_{ij}^{(\alpha \beta)}(a_{\beta} \omega) = a_{\alpha}a_{\beta}
		\int_{0}^{\infty}ds~ e^{i \omega a_{\beta} s} G^{(\alpha \beta)}_{ij}(s).
	\end{equation}
Notice that we have moved the upper limit of integration over $ds$ to infinity as $\tau_{\cal B} \ll \Delta \eta$ and the lower limit of the integral over $d\eta_{1}$ was simply taken to be $\eta$ since only small values of $s$ contribute significantly to the integral. 

In the following we will study the form of the function $J(x)$. We can solve explicitly the integral:
	\begin{align}
	J(\omega^{\prime} a_{\alpha} - \omega a_{\beta}) &= \int_{\eta}^{\eta + \Delta \eta}d\eta_{1}~\frac{e^{i(\omega^{\prime} a_{\alpha} - \omega a_{\beta})\eta_{1}}}{\Delta \eta}\nonumber\\
	 &= e^{i(\omega^{\prime} a_{\alpha} - \omega a_{\beta})\eta} F\left(\omega^{\prime} a_{\alpha} - \omega a_{\beta}\right),
	\end{align}
with
	\begin{equation}
	F\left(\omega^{\prime} a_{\alpha} - \omega a_{\beta}\right) = e^{i(\omega^{\prime} a_{\alpha} - \omega a_{\beta})\frac{\eta}{2}}\frac{\sin \left[\left(\frac{\omega^{\prime} a_{\alpha} - \omega a_{\beta}}{2}\right)\Delta \eta\right]}{\left[\left(\frac{\omega^{\prime} a_{\alpha} - \omega a_{\beta}}{2}\right)\Delta \eta\right]}.
	\end{equation}														 	
We observe that, for $|\omega^{\prime} a_{\alpha} - \omega a_{\beta}| \ll (\Delta \eta)^{-1}$, the function 
$|F|$ is very close to one, and, on the other hand, for $|\omega^{\prime} a_{\alpha} - \omega a_{\beta}| \gg (\Delta \eta)^{-1}$, $|F|$ is very close to zero. In other words, $|F|$ is sharply peaked around the value 
$|\omega^{\prime} a_{\alpha} - \omega a_{\beta}| = 0$. Therefore, within our assumptions, we can make the approximation
	\begin{equation}\label{eqq:J_delta}
	J(\omega^{\prime} a_{\alpha} - \omega a_{\beta}) \approx   e^{i(\omega^{\prime} a_{\alpha} - \omega a_{\beta})\eta} \delta\left(\omega^{\prime} a_{\alpha} - \omega a_{\beta}\right).
	\end{equation}
Let us now obtain the standard form of the master equation. By {\it standard form} we mean the derivation known by the Lindblad-Kossakowski equation \cite{lindblad1976generators,gorini1976completely}. We start by inserting Eq. \eqref{eqq:J_delta} into equation \eqref{eqq:gme_nochange_3}. We obtain
	\begin{align}
	\frac{\Delta \tilde{\rho}_{A}(\eta)}{\Delta \eta} = \sum_{\omega}\sum_{i,j}\sum_{\alpha, \beta}  W_{ij}^{(\alpha \beta)}(a_{\beta} \omega)\left( A^{(\beta)}_{j}(\omega)\tilde{\rho}_{A}(\eta) A^{\dagger (\alpha)}_{i}\left(\frac{a_{\beta}}{a_{\alpha}} \omega\right)-A^{\dagger (\alpha)}_{i}\left(\frac{a_{\beta}}{a_{\alpha}} \omega\right) A^{(\beta)}_{j}(\omega)\tilde{\rho}_{A}(\eta)\right)\nonumber\\
	+ \sum_{\omega}\sum_{i,j}\sum_{\alpha, \beta}  W_{ij}^{*(\alpha \beta)}(a_{\alpha} \omega)\left( A^{ (\alpha)}_{i}(\omega)\tilde{\rho}_{A}(\eta) A^{\dagger (\beta)}_{j}\left(\frac{a_{\alpha} \omega}{a_{\beta}}\right)-\tilde{\rho}_{A}(\eta)A^{\dagger(\beta)}_{j}\left(\frac{a_{\alpha} \omega}{a_{\beta}}\right)A^{ (\alpha)}_{i}(\omega) \right) .  \label{eqq:gme_stanform_1}  
	\end{align}
This equations describes the change of $\tilde{\rho}_{A}(\eta)$ within the interval $(\eta, \eta + \Delta\eta)$. The ratio $\Delta \tilde{\rho}_{A}/\Delta \eta$ can be thought as an operation of averaging which smooths out rapid variations of $\tilde{\rho}_{A}(\eta)$. The investigation of the time evolution on a time scale 
$\tau_{\cal A} \ll \Delta\eta$ allows us to neglect these small and rapid fluctuations. This coarse-graining approximation is also known as the {\it Markovian approximation}. Moreover, in our case the operators $A(\omega)$ will take the form of linear combinations of the Pauli matrices. Then
$\omega$ is to be associated with the eigenvalues of the atomic Hamiltonian $(\mathcal{H}_{A})$. Therefore, we can write $A(\lambda \omega) = A(\omega)$ for some positive constant $\lambda$. Hence 
	\begin{align}
	&\frac{d}{d \eta}\tilde{\rho}_{A}(\eta) = \sum_{\omega}\sum_{i,j}\sum_{\alpha, \beta}  W_{ij}^{(\alpha \beta)}(a_{\beta} \omega)
	\left( A^{(\beta)}_{j}(\omega)\tilde{\rho}_{A}(\eta) A^{\dagger (\alpha)}_{i}(\omega)-A^{\dagger (\alpha)}_{i}(\omega) A^{(\beta)}_{j}(\omega)\tilde{\rho}_{A}(\eta)\right)\nonumber\\
	+&\sum_{\omega}\sum_{i,j}\sum_{\alpha, \beta}  W_{ji}^{*(\beta \alpha)}(a_{\beta} \omega)
	\left( A^{ (\beta)}_{j}(\omega)\tilde{\rho}_{A}(\eta) A^{\dagger (\alpha)}_{i}(\omega)-\tilde{\rho}_{A}(\eta)A^{\dagger(\alpha)}_{i}(\omega)A^{ (\beta)}_{j}(\omega) \right).\label{eqq:gme_stanform_ji}  
	\end{align}
where we have changed $i \longleftrightarrow j$ and $\alpha \longleftrightarrow \beta$ in the second term of the right-hand side. Moving to the Schr\"odinger representation, 
$\tilde{\rho}_{A}(\eta) = e^{i \mathcal{H}_{A} \eta}\rho_{A}(\eta)e^{-i \mathcal{H}_{A} \eta}$, and also
	\begin{equation}
	e^{i \mathcal{H}_{A} \eta} A_{i}^{(\alpha)}(\omega) e^{-i \mathcal{H}_{A} \eta} = e^{- i \omega \eta}  A_{i}^{(\alpha)}(\omega),
	\end{equation}
and defining the following functions
	\begin{align}
	\Gamma^{(\alpha \beta)}_{ij}(a_{\beta}, \omega) =  W_{ij}^{(\alpha \beta)}(a_{\beta} \omega) +  W_{ji}^{*(\beta \alpha)}(a_{\beta} \omega),\label{eqq:gamma_gen}\\
	\Delta^{(\alpha \beta)}_{ij}(a_{\beta}, \omega) = \frac{W_{ij}^{(\alpha \beta)}(a_{\beta} \omega) -  W_{ji}^{*(\beta \alpha)}(a_{\beta} \omega)}{2i}, \label{eqq:Delta_gen}
\end{align}	
we obtain that
	\begin{align}
&\dot{\rho}_{A}(\eta) = - i \left[\mathcal{H}_{A},\rho_{A}(\eta)\right]
 - i \sum_{\omega}\sum_{i,j}\sum_{\alpha, \beta}\Delta^{(\alpha \beta)}_{ij}(a_{\beta}, \omega)\left[A^{\dagger (\alpha)}_{i}(\omega) A^{(\beta)}_{j}(\omega),\rho_{A}(\eta)\right]  \nonumber\\
&+\frac{1}{2}\sum_{\omega}\sum_{i,j}\sum_{\alpha, \beta} \Gamma^{(\alpha \beta)}_{ij}(a_{\beta}, \omega)\left( 2 A^{(\beta)}_{j}(\omega)\rho_{A}(\eta) A^{\dagger (\alpha)}_{i}(\omega)
-\left\lbrace A^{\dagger (\alpha)}_{i}(\omega) A^{(\beta)}_{j}(\omega),\rho_{A}(\eta)\right\rbrace \right),\label{eqq:gme_stanform_complete}
	\end{align}
where $\dot{F} = dF/d\eta$ and the commutator and the anticommutator of operators $C,D$ are defined in the usual way. We define the following effective Hamiltonian
	\begin{equation}\label{eqq:h_eff}
	\mathcal{H}_{\textrm{eff}} \equiv \mathcal{H}_{A} + \sum_{\omega}\sum_{i,j}\sum_{\alpha, \beta}\Delta^{(\alpha \beta)}_{ij}(a_{\beta}, \omega)A^{\dagger (\alpha)}_{i}(\omega) A^{(\beta)}_{j}(\omega),
	\end{equation}
which describes the unitary evolution of the subsystem. The term $\Delta^{(\alpha \beta)}_{ij}(a_{\beta}, \omega)$ produces a divergent contribution in $\mathcal{H}_{\textrm{eff}}$ -- this is just the well known Lamb shift, which is also expected to appear due to the coupling with the fields. The emergence of such divergences is associated with the nonrelativistic treatment of the dynamics of subsystem $A$ -- a renormalization procedure is therefore required~\cite{benatti04_pra}. However, observe that this contribution only has an impact on the unitary evolution. Since in the present study we are only interested in the dynamics of entanglement, which is not affected by the Lamb shift, we shall not discuss it further. In any case, we emphasize the importance of an adequate treatment of the Lamb contribution in the analysis of the unitary evolution of the subsystem $A$. A nice discussion on this topic can be found in Ref.~\cite{benatti04_pra}.

From here, we obtain the standard form of the generalized master equation  describing the time evolution of atoms travelling along different world lines:
	\begin{equation}\label{eqq:gme_final_form}
	\dot{\rho}_{A}(\eta) = - i \left[\mathcal{H}_{\textrm{eff}},\rho_{A}(\eta)\right] + \mathcal{L}\left\lbrace\rho_{A}(\eta)\right\rbrace,
	\end{equation}	
where the term 	$\mathcal{L}\left\lbrace\rho_{A}(\eta)\right\rbrace$, the dissipative contribution to generalized master equation , is given by
	\begin{equation}
	\mathcal{L}\left\lbrace\rho_{A}(\eta)\right\rbrace = \frac{1}{2}\sum_{\omega}\sum_{i,j}\sum_{\alpha, \beta} \Gamma^{(\alpha \beta)}_{ij}(a_{\beta}, \omega)
	\left( 2 A^{(\beta)}_{j}(\omega)\rho_{A}(\eta) A^{\dagger (\alpha)}_{i}(\omega)
	-\left\lbrace A^{\dagger (\alpha)}_{i}(\omega) A^{(\beta)}_{j}(\omega),\rho_{A}(\eta)\right\rbrace\right).\label{eqq:dissipative_term}
	\end{equation}
As we shall see, it is the non-unitary term 	$\mathcal{L}\left\lbrace\rho_{A}(\eta)\right\rbrace$ that is responsible for decoherence effects as well as entanglement generation. Observe the generality of our derivation of the generalized master equation, that can be applied to a scalar or a vector field.

\section{Generalized master equation for a couple of two-level systems}\label{sec:thegme_coupleofatoms}

Our task now is to specify the generalized master equation for one particular case. The generalization of the results presented in Ref.~\cite{benatti04_pra} will be obtained. We consider the situation of two identical two-level systems interacting with a quantum massless scalar field $\phi(x)$. The parameter we choose to describe the time evolution of the system is the proper time $(\tau)$ of one of the two-level systems. The different contributions to the Hamiltonian of the system can be cast in the following form
\begin{equation}
\mathcal{H} = \mathcal{H}_A + \mathcal{H}_F + \mathcal{H}_{int} 
\end{equation}
where
	\begin{align}
		&\mathcal{H}_{A} = \frac{\omega_{0}}{2}\left[S^{z}_{(1)}\otimes\unit_{2} \frac{d\tau_{1}(\tau)}{d\tau} + \unit_{1}\otimes S_{(2)}^{z}\frac{d\tau_{2}(\tau)}{d\tau} \right],\\
		&\mathcal{H}_{F} = \frac{1}{2}\int d^3\textbf{x} \left[(\dot{\phi}(x))^2 +(\nabla \phi(x))^2 \right],\\
		&\mathcal{H}_{int} =  \sum_{j = 1}^{2}\mu_{j}m^{(j)}(\tau_{j}(\tau)) \phi\left[x_{j}(\tau_{j}(\tau))\right]\frac{d\tau_{j}(\tau)}{d\tau} .
		\label{eqq:inthamilt_scalar} 
	\end{align}
Raising and lowering off-diagonal operators $S_{(j)}^{\pm}$ which act in the $j-$th two-level systems subspaces are given by
	\begin{equation}
	S_{(j)}^{+} = \ket{e_{j}}\bra{g_{j}}; \quad S_{(j)}^{-} = \ket{g_{j}}\bra{e_{j}} .
	\end{equation}
The diagonal operator $S_{(j)}^{z}$ reads
	\begin{equation}
		S^{z}_{(j)} = \frac{1}{2}\left(\ket{e_{j}}\bra{e_{j}} - \ket{g_{j}}\bra{g_{j}}\right).
	\end{equation}
In addition, $\mu_{j}$ is the coupling constant describing the interaction of the $j-$th two-level system with the quantum field and $m^{(j)}$ is the monopole moment of the two-level systems, given by
	\begin{align}
	m^{(j)}(\tau_{j}(\tau)) =  S^{+}_{(j)}e^{i \omega_0 \tau_{j}(\tau)} + S^{-}_{(j)}e^{-i \omega_0 \tau_{j}(\tau)}.\label{eqq:monopole_moment}
	\end{align}	
For simplicity, henceforth we take $\mu_1 = \mu_2 = \mu$. We would like to study the situation where both two-level systems are in uniformly accelerated world lines. It is known that the coordinates of these two-level systems are given by the so called \textit{Rindler coordinates}. These coordinates for uniform acceleration occurring only in the $x-$direction are expressed as
	\begin{equation}
	t  = \frac{e^{a \xi}}{a} \sinh \left( a \eta \right), 
	\quad x = \frac{e^{a \xi}}{a} \cosh \left(a \eta \right).
	\label{Rindler}
	\end{equation}
The proper acceleration is defined by $\alpha^{-1} = a e^{-a\xi_{i}}$ and the proper time is associated with 
$\eta$ and $\xi$ via the relation $\tau = e^{a\xi} \eta$. Note that this parameter $\eta$ is not necessarily the same $\eta$ of the previous section, even though in principle one is free to make this choice. We consider that our two-level systems are traveling along different hyperbolic trajectories 
$x^{\mu}(\tau_1) = (t(\tau_1),x(\tau_1),y_1,z_1)$ and $x^{\mu}(\tau_2) = (t(\tau_2),x(\tau_2),y_2,z_2)$, whose functional dependence $t(\tau_i), x(\tau_i)$ is given by equations similar to Eq.~(\ref{Rindler}).

We now identify the time-evolution parameter of the generalized master equation derived in the previous section with the proper time of one of the two-level systems. This allows us to perform a simple change of variables in the generalized master equation. However, recall that we have two different proper times at our disposal. Nevertheless, considering that the two different hyperbolas are in the same Rindler wedge, they are cut by a single line of constant Rindler coordinate $\eta$, which means that we have a simple relation between the proper times of the two-level systems. In what follows we find it convenient to choose the proper time $\tau_1$ as the parameter that describes the time evolution of the system. Using standard methods, we can then relate the other proper time $\tau_2$ to $\tau_1$. This relation is given by 
	\begin{equation}\label{eqq:parametrization_t2_t1}
	\tau_{2}(\tau_{1}) = \tau_{1}e^{a(\xi_2 - \xi_1)} = \frac{\alpha_{2}}{\alpha_{1}}\tau_{1},
	\end{equation}		
and, therefore, the $a_{\beta}$ constants that we have used in Eq. \eqref{eqq:a-deff} can be written as $a_{1} = 1$ and $a_{2} = \alpha_2 /\alpha_1$. Observe that, in this situation of equal Rindler times (but different hyperbole and, as a result, different proper times), the two two-level systems are always separated by a spacelike distance, even though they lie in the same Rindler wedge.

On the other hand, as we are dealing with a scalar field, the $i,j-$summation is removed. In addition, the $A$ operators are to be identified with the ladder operators $S^{\pm}$. We then write the generalized master equation  as
	\begin{equation}\label{eqq:gme_scalar_field}
	\frac{d\rho_{A}(\tau)}{d\tau} = - i \left[\mathcal{H}_{\textrm{eff}},\rho_{A}(\tau)\right] + \mathcal{L}\left\lbrace\rho_{A}(\tau)\right\rbrace,
	\end{equation}	
with $\tau \equiv \tau_1$, where the dissipative contribution is given by
	\begin{align}
	 &\mathcal{L}\left\lbrace\rho_{A}(\tau)\right\rbrace = \frac{1}{2}\sum_{\alpha, \beta = 1}^{2} \Gamma^{(\alpha \beta)}(a_{\beta}\omega)\nonumber\\
	 &\times \left( 2 S_{(\beta)}^{-}\rho_{A}(\tau) S_{(\alpha)}^{+}-\left\lbrace S_{(\alpha)}^{+}S_{(\beta)}^{-},\rho_{A}(\tau)  \right\rbrace\right) + (\omega \rightarrow -\omega),\label{eqq:dissipative_term_scalar}
	\end{align}
with $\mathcal{H}_{\textrm{eff}}$ being given by
	\begin{equation}\label{eqq:h_eff_scalar}
	\mathcal{H}_{\textrm{eff}} = \mathcal{H}_{A} + \sum_{\alpha, \beta}\Delta^{(\alpha \beta)}(a_{\beta} \omega) S_{(\alpha)}^{+}S_{(\beta)}^{-} + (\omega \rightarrow -\omega).
	\end{equation}
The $\Gamma(x)$ and $\Delta(x)$ functions are given by Eqs. \eqref{eqq:gamma_gen} and \eqref{eqq:Delta_gen}, removing the dependence on the $i$ and $j$ indices. We find that
	\begin{equation}\label{eqq:w_scalar_function}
	W^{(\alpha \beta)}(a_{\beta} \omega) = a_{\alpha}a_{\beta} \int_{0}^{\infty}ds~e^{i \omega a_{\beta} s} G^{+(\alpha \beta)}(s),
	\end{equation}
where $G^{+(\alpha \beta)}(s)$ is the positive-frequency Wightman function whose form will be presented in the following. 

In order to compute $\Gamma^{(\alpha \beta)}( a_{\beta}\omega)$, we must compute first the functions $G^{+ (\alpha \beta)}(s)$ and, for consequence, $W^{(\alpha \beta)}(a_{\beta} \omega)$. The positive-frequency Wightman function for a massless real scalar field in Minkowski spacetime is given by 
	\begin{equation}\label{eqq:wight_inertial}
	G^{+}(x,x')	= - \frac{1}{4 \pi^2 }\frac{1}{\left[(t - t' - i \epsilon)^2 - | \textbf{x} - \textbf{x}'|^2\right]}.
	\end{equation}
As we require that the atoms are in a uniformly accelerated motion coupled with a scalar field prepared in the Minkowski vacuum $\ket{0,M}$, the procedure is to change the inertial coordinates $(x,x')$ in Eq. \eqref{eqq:wight_inertial} to the Rindler coordinates using the expressions defined above~\cite{Rodriguez-Camargo:2016fbq}. By doing so we have, for the usual term represented by $\alpha = \beta$
	\begin{equation}\label{eqq:PWF_II}
	G^{+(\beta \beta)}(s) = - \frac{1}{16 \pi^2 \alpha_{\beta}^2} \frac{1}
	{\sinh^2 \left(\frac{1}{2 \alpha_1}(s - 2 i \epsilon)\right)},
	\end{equation}
and for the crossed terms $\alpha \neq \beta$ we have $G^{+ (12)}(s) = G^{+ (21)}(s) = G^{+}_{c}(s)$, where
	\begin{align}
	  G^{+}_{c}(s) = - \frac{1}{16 \pi^2 \alpha_1 \alpha_2} \left[\sinh \left(\frac{s}{2 \alpha_1} - \frac{4 i \epsilon}{\alpha_1 + \alpha_2 } + \frac{\phi}{2}\right)\right.\nonumber\\
	  \times\left.\sinh \left(\frac{s}{2 \alpha_1} - \frac{4 i \epsilon}{\alpha_1 + \alpha_2 } - \frac{\phi}{2}\right)\right]^{-1},\label{eqq:PWF_IJ}
	\end{align}	
with the following definition
	\begin{equation}\label{coshphi}
	\cosh \phi = 1 + \frac{(\alpha_1 - \alpha_2 )^2 + |\Delta \textbf{y}|^2 }{2 \alpha_1 \alpha_2},
\end{equation}	
where $|\Delta \textbf{y}|^2 = (y_2 - y_1)^2 + (z_1 - z_2)^2$ is the orthogonal (to the $(t,x)-$plane) distance between the two-level systems which we will assume to be constant. Observe that the regular $\epsilon$ must be kept finite so that physical input on the two-level systems are not incompatible with the usual conditions imposed on the Wightman distribution function. It is only in this case that the limit of sharp switching is well defined -- it holds as a valid approximation for fixed energy gaps when the proper time interval is much longer than the $\epsilon$. For a thorough discussion of this point, see Ref.~\cite{Rodriguez-Camargo:2016fbq} and references cited therein.

Let us calculate the different contributions to the dynamics separately. Observe that
	\begin{align}
		\Gamma^{(\alpha \beta)}(a_{\beta} \omega) &= W^{(\alpha \beta)}(a_{\beta} \omega) + W^{* (\beta \alpha)}(a_{\beta} \omega)\nonumber \\
		&= a_{\alpha}a_{\beta}\int_{0}^{\infty}ds e^{i \omega a_{\beta} s} G^{+(\alpha \beta)}(s)\nonumber\\
		 &+ a_{\beta}a_{\alpha}\int_{0}^{\infty}ds e^{-i \omega a_{\beta} s} G^{+(\beta \alpha)}(s).\label{eqq:gamma_ii_explicit}
	\end{align}
By changing $s \rightarrow -s$ in the second integral and observing that $G^{+(\beta \alpha)}(-s) = G^{+(\alpha \beta)}(s)$ we have that
	\begin{equation}\label{eqq:gamma_ii_fourier}
	\Gamma^{(\alpha \beta)}(a_{\beta}  \omega) = a_{\alpha}a_{\beta} \int_{- \infty}^{\infty}ds e^{i \omega a_{\beta} s} G^{+(\alpha \beta)}(s).
	\end{equation}
The explicit computation of Eq. \eqref{eqq:gamma_ii_fourier} is given in the Appendix \ref{app:gammas}. We find that
	\begin{align}
	\Gamma^{(\beta \beta)}(a_{\beta}  \omega) = \Gamma^{(\beta \beta)}(\omega)  = \frac{\omega \alpha_{\beta}}{2 \pi \alpha_{1}}\left(\frac{1}{1 - e^{- 2\pi \omega \alpha_{\beta}}}\right),\label{eqq:gamma_ii}
	\end{align}
and for the crossed contribution
	\begin{align}
	\Gamma^{(\alpha \beta )}(a_{\beta} \omega) &= 	\Gamma^{(\alpha \beta)}(\omega)\nonumber\\
	&=  \frac{\sin \left(\omega \alpha_{\beta}\phi\right)}{2 \pi \alpha_{1} \sinh \phi}\left(\frac{1}{1 - e^{- 2\pi \omega \alpha_{\beta}}}\right).\label{eqq:gamma_ij}
	\end{align}
It is useful to write Eq. \eqref{eqq:gme_scalar_field} in terms of the pseudo-spin operators using that $S_{(\alpha)}^{+} = S_{(\alpha)}^{1} + i S_{(\alpha)}^{2}$ and $S_{(\alpha)}^{-} = S_{(\alpha)}^{1} - i S_{(\alpha)}^{2}$, where $S_{(1)}^{i} = \sigma^{i}\otimes\sigma^{0}$ and $S_{(2)}^{i} = \sigma^{0}\otimes\sigma^{i}$ with $\sigma^{i}$ the $i$-th Pauli matrix. See, for example, Ref. \cite{FICEK2002369}. The dissipative contribution can be written as (hereafter we omit the subscript $A$ in the reduced density matrix $\rho_{A}$)
	\begin{align}
	\mathcal{L}\left\lbrace\rho(\tau)\right\rbrace = \frac{1}{2} \sum_{i,j = 1}^{3}\sum_{\alpha, \beta = 1}^{2} \mathcal{C}_{ij}^{(\alpha \beta)}\left[2 S^{j}_{(\beta)} \rho(\tau) S^{i}_{(\alpha)}
	\right.\nonumber\\
	 - \left.\left\lbrace S^{i}_{(\alpha)}S^{j}_{(\beta)}, \rho(\tau) \right\rbrace\right],\label{eqq:dissi_kossakowski}
	\end{align}
with the Kossakowski matrix $ \mathcal{C}_{ij}^{(\alpha \beta)}$ defined as
	\begin{equation}\label{eqq:kossakow_matrix}
	 \mathcal{C}_{ij}^{(\alpha \beta)} = A^{(\alpha \beta)} \delta_{ij} - i B^{(\alpha \beta)} \epsilon_{ijk}\delta_{3k} -  A^{(\alpha \beta)}\delta_{3i}\delta_{3j}.
	\end{equation}	 	
We have also defined the following functions
	\begin{align}
	A^{(\alpha \beta)} = \frac{\mu^{2}}{4}\left(\Gamma^{(\alpha \beta)}( \omega) + \Gamma^{(\alpha \beta)}(-  \omega) \right),\label{eqq:kossa_A_general}\\
	B^{(\alpha \beta)} = \frac{\mu^{2}}{4}\left(\Gamma^{(\alpha \beta)}(  \omega) - \Gamma^{(\alpha \beta)}(-     \omega) \right).\label{eqq:kossa_B_general}
	\end{align}
Using Eqs. \eqref{eqq:gamma_ii} and \eqref{eqq:gamma_ij} we can compute explicitly the functions given by Eqs. \eqref{eqq:kossa_A_general} and \eqref{eqq:kossa_B_general}. For $\alpha = \beta$ we have
	\begin{align}
	&A^{(\beta)} = A^{(\beta \beta)} =  \Gamma_{0}\frac{ \alpha_{\beta}}{4\alpha_{1}}\left(1 + \frac{2}{e^{2 \pi \alpha_{\beta} \omega} - 1}\right),\label{eqq:a_ii_solved}\\
	&B^{(\beta)} = B^{(\beta \beta)} = \Gamma_{0}\frac{ \alpha_{\beta}}{4\alpha_{1}},\label{eqq:b_ii_solved}
	\end{align}		
and for $\alpha \neq \beta$,
\begin{align}
&A^{(\alpha \beta)} = \Gamma_{0}\frac{\sin(\omega \alpha_{\beta} \phi)}{4 \omega \alpha_{1} \sinh \phi} \left(1 + \frac{2}{e^{2 \pi \alpha_{\beta} \omega} - 1}\right),\label{eqq:aij_final}\\
&B^{(\alpha \beta)} = \Gamma_{0}\frac{\sin(\omega \alpha_{\beta} \phi)}{4 \omega \alpha_{1} \sinh \phi}, \label{eqq:bij_final}
\end{align}	
where we have introduced $\Gamma_{0} = \mu^{2} \omega / 2 \pi$ being the spontaneous emission rate.

\section{Entanglement dynamics}\label{sec:ent_dynamics}

To describe entanglement dynamics we may ignore the unitary contribution given by Eq.~\eqref{eqq:h_eff_scalar}. We write the dissipative part given by Eq.~\eqref{eqq:dissi_kossakowski} in the basis $\left\lbrace\ket{1} = \ket{g_1}\otimes\ket{g_2}, \ket{2} = \ket{e_1}\otimes\ket{e_2},\right.$ $\left.\ket{3} =\ket{g_1}\otimes\ket{e_2}, \ket{4} = \ket{e_1}\otimes\ket{g_2}  \right\rbrace$. It can be proved that if we assume a block-diagonal form of the density matrix for $\tau=0$ the evolution of  $\rho(\tau)$ does not change this block-diagonal form. Therefore, we assume that the density matrix in the aforementioned basis has the form
	\begin{equation}\label{eqq:rho_initial}
	\rho(\tau) = \left(\begin{array}{cccc}
	\rho_{11} & \rho_{12} & 0 & 0\\
	\rho_{21} & \rho_{22} & 0 & 0\\
	0 & 0 & \rho_{33} & \rho_{34}\\
	0 & 0 & \rho_{43} & \rho_{44}
	\end{array}\right).
	\end{equation}
The problem simplifies by working in the coupled basis of the system. For identical two-level systems, such a basis reads 
	\begin{align}
	\ket{G} &= \ket{g_1}\otimes\ket{g_2},\label{basis:1}\\
	\ket{E} &= \ket{e_1}\otimes\ket{e_2},\label{basis:2}\\
	\ket{S} &= \frac{1}{\sqrt{2}}(\ket{g_1}\otimes\ket{e_2} + \ket{e_1}\otimes\ket{g_2}),\label{basis:3}\\
	\ket{A} &= \frac{1}{\sqrt{2}}(\ket{e_1}\otimes\ket{g_2} - \ket{g_1}\otimes\ket{e_2}),\label{basis:4}
	\end{align}
and we obtain that in this basis, the density matrix still has a block diagonal form as
	\begin{equation}\label{eqq:rho_initial_basis}
	\rho(\tau) = \left(\begin{array}{cccc}
	\rho_{GG} & \rho_{GE} & 0 & 0\\
	\rho_{EG} & \rho_{EE} & 0 & 0\\
	0 & 0 & \rho_{SS} & \rho_{SA}\\
	0 & 0 & \rho_{AS} & \rho_{AA}
	\end{array}\right),
	\end{equation}
with $\rho_{ij} = \bra{i}\rho\ket{j}$. We compute the matrix elements of the generalized master equation given by Eq. \eqref{eqq:gme_scalar_field} in the aforementioned basis. We obtain eight coupled linear differential equations involving all the matrix components of Eq. \eqref{eqq:rho_initial_basis}. As we are dealing with a more general situation than the ones studied in Refs.~\cite{benatti04_pra, yu2016pra}, our set of equations is more involved and are presented in the appendix~\ref{app:gmeterms}. We observe from Eqs.~(\ref{appeq-gme1}-\ref{appeq-gme8}) that if we take $\alpha_1 = \alpha_2$ ,$i.e.$, set both two-level systems at the same world line, we then have $A^{(11)} = A^{(22)}$, $B^{(11)} = B^{(22)}$, $A^{(12)} = A^{(21)}$ and $B^{(12)} = B^{(21)}$. This simplifies the set of equations and we recover the previous result of the literature \cite{benatti04_pra,hu2015, yu2016pra}. Therefore, as expected, the case where the two-level systems are with the same proper acceleration is a special case of the master equation derived in this work.

The problem follows by treating the two two-level system as a single four-level system. Next, we must study  how much of entanglement is stored in this quantum system. There are many alternatives to proceed this study. We here choose to compute the \textit{concurrence} of this system. The concurrence $C(\rho)$ is a function of the eingenvalues of the matrix $R = \rho \tilde{\rho}$, with $\tilde{\rho}$ given by
	\begin{equation}
	\tilde{\rho} = \sigma_{y}\otimes \sigma_{y} \rho^{T} \sigma_{y}\otimes \sigma_{y},
	\end{equation}	
which has value $C =1$	for maximally entangled state and $C = 0$ for separable states. In the basis given by Eqs. \eqref{basis:1}, \eqref{basis:2}, \eqref{basis:3} and \eqref{basis:4} we have the following form of the concurrence \cite{yu2016pra, Tana_2004}
	\begin{equation}\label{eqq:concurrence}
	C[\rho(\tau)] = \textrm{max} \left\lbrace0, \mathcal{C}_{1}(\tau),\mathcal{C}_{2}(\tau)\right\rbrace,
	\end{equation}
with
	\begin{align}
	\mathcal{C}_{1}(\tau) = &\sqrt{\left[\rho_{AA}(\tau) - \rho_{SS}(\tau) \right]^2 - \left[\rho_{AS}(\tau) - \rho_{SA}(\tau)\right]^2}\nonumber\\
	&- 2 \sqrt{\rho_{GG}(\tau) \rho_{EE}(\tau)},\\
	\mathcal{C}_{2}(\tau) = &- \sqrt{\left[\rho_{AA}(\tau) + \rho_{SS}(\tau) \right]^2 - \left[\rho_{AS}(\tau) + \rho_
{SA}(\tau)\right]^2}\nonumber\\
	&+ 2|\rho_{GE}(\tau)|.
	\end{align}	
In the following, we numerically solve the dissipative part of the generalized master equation for different initial configuration of the two-level systems and use this solution to evaluate the concurrence of Eq. \eqref{eqq:concurrence} to study the entanglement dynamics of this system.

\subsection{Entanglement degradation}

Let us first consider the system to be initially in a maximally entangled state and compute its evolution. It is well known that this setup provides the scenario to study a phenomenon called \textit{entanglement degradation}. The interest in investigating this lies in the fact that entanglement degradation for accelerated atoms will behave differently from inertial atoms due to the Unruh effect \cite{gallock2021entangled}. This phenomenon can be investigated by using as the initial state the symmetric state $\ket{S}$ or the antisymmetric state $\ket{A}$. It is important to point out that many works consider a non-trivial orthogonal distance $|\Delta \textbf{y}|$ between the two-level systems~\cite{hu2015,yu2016pra}. In the present study, this effect can be observed as a change in the $\phi$ function defined by Eq. \eqref{coshphi}. We observed that this modification leads to a change in the concurrence that is irrelevant compared with the change in the proper acceleration of the two-level systems. By this reason we choose to work with $|\Delta \textbf{y}|^2 = 0$.

\begin{figure}[H]
\centering
\includegraphics[scale=0.75]{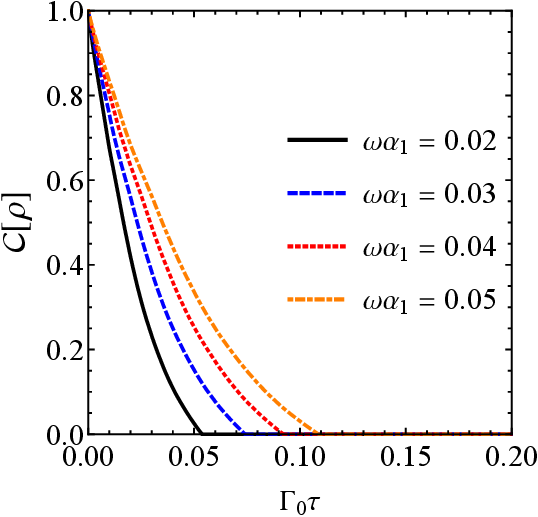}
\caption{Comparison of entanglement degradation for different values of $\omega \alpha_1$. The initial state was prepared to be $\ket{S}$. We use  $\omega \alpha_2 = 0.6$.} 
\label{fig:ent_death_s}
\end{figure}

\begin{figure}[H]
\centering
\includegraphics[scale=0.75]{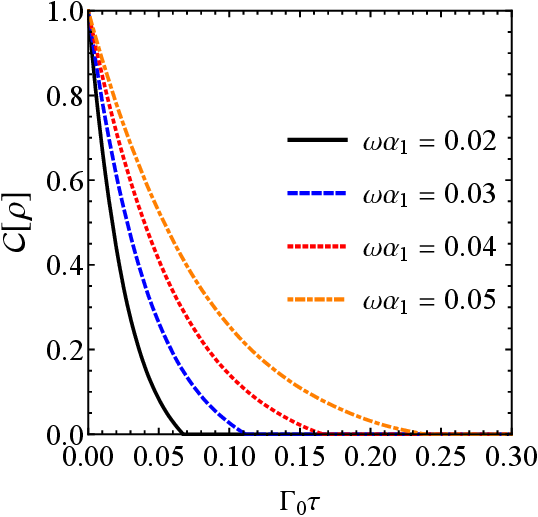}
\caption{Comparison of entanglement degradation for different values of $\omega \alpha_1$. The initial state was prepared to be $\ket{A}$. We use $\omega \alpha_2 = 0.6$.} 
\label{fig:ent_death_a}
\end{figure}

For the the system prepared in the symmetric state the results are depicted in Fig. \ref{fig:ent_death_s}, where we have fixed the value of acceleration of the second two-level system  $(\alpha_2^{-1})$ and varied the acceleration of the first two-level system $(\alpha_1^{-1})$. In Fig. \ref{fig:ent_death_a} we used the antisymmetric state as the initial state of the system where $\alpha_2$ is fixed and we varied $\alpha_1$. We observe that the entanglement degradation occurs faster for the system prepared in the state $\ket{S}$. 
Moreover, notice the early-stage disentanglement developed by the two-level system -- the profile seen in the degradation of the correlations of the two-level systems by environmental noise is compatible with the known effect of entanglement sudden death~\cite{doi:10.1126/science.1167343}. Such results are in agreement with the outcomes presented in Ref.~\cite{hu2015}.

\begin{figure}[H]
\centering
\includegraphics[scale=0.7]{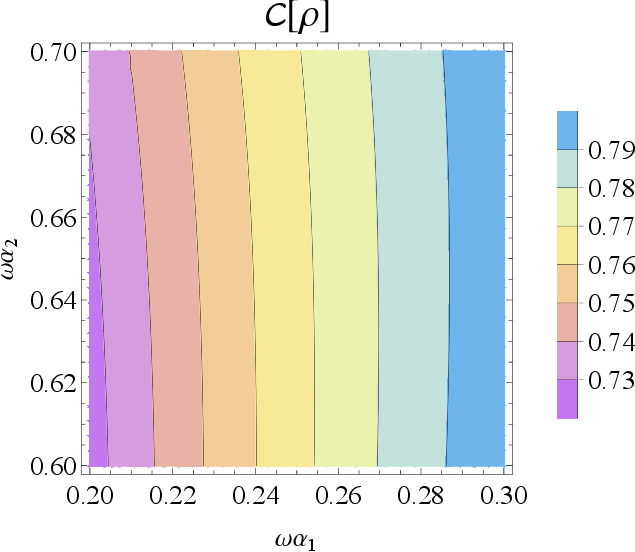}
\caption{Comparison of entanglement degradation for different values of $\omega \alpha_1$ and $\omega \alpha_2$. The initial state was prepared to be $\ket{S}$ and we fixed the parameter $\Gamma_{0}\tau=0.05$.} 
\label{fig:ent_death_accels}
\end{figure}

The master equation derived in this work allows us to investigate how the entanglement occurs when the two-level systems are in distinct world lines, $i.e.$, for different values of $\alpha_{i}$. Here we consider $\alpha_1 \neq \alpha_2$ but observe that for $\alpha_1 = \alpha_2$ the master equation is the same as the one used in Ref. \cite{benatti04_pra} and, therefore, we obtain the same results. By construction we have $\alpha_2 \geq \alpha_1$ and we discuss how entanglement behaves when we vary both proper accelerations. In Fig. \ref{fig:ent_death_accels} we have for a fixed value of time parameter $\Gamma_{0}\tau$ the values that the concurrence $C(\tau)$ assumes for different values of $\alpha_1$ and $\alpha_2$. We observe that as both accelerations decrease (as $\alpha_{i}$ increases) the lifetime of the entangled state increases.

\subsection{Entanglement Harvesting}

\begin{figure}[H]
\centering
\includegraphics[scale=0.7]{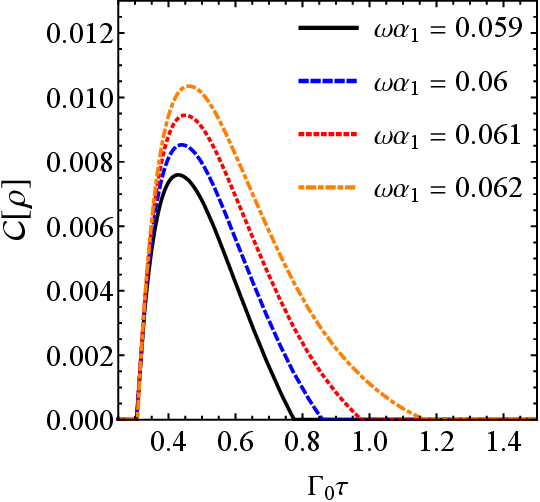}
\caption{Comparison of entanglement harvesting for different values of $\omega \alpha_1$. The initial state was prepared to be $\ket{G}$. We use $\omega \alpha_2 = 0.58$.} 
\label{fig:ent_harv_g}
\end{figure}

Another interesting phenomenon is the so called \textit{entanglement harvesting} \cite{martinmartinez-PhysRevD.92.064042}. This phenomenon occurs when a system is prepared in a separable state and, due to the interaction of the atoms with the field, an atomic entangled state is created. As discussed above, for a massless scalar field, the system is truly harvesting entanglement from the field only if the two-level systems are separated by a spacelike interval \cite{martinmartinez_prd21}. Recall that, in our system we have a pair of two-level systems traveling along different hyperbole, i.e,  with different proper accelerations, but along lines of constant Rindler time. Hence they are indeed separated by a spacelike distance. In this scenario we can definitely assert that the entanglement that is being generated has to come from entanglement harvesting. To study this phenomena within our formalism we prepare the two-level system in the $\ket{G}$ state or in the $\ket{E}$ state. We observe in Fig. \ref{fig:ent_harv_g} that, for the initial state being $\ket{G}$, an entangled state can be created and the concurrence increases with the value of $\alpha_1$. One interesting result is that entanglement is only created by smaller values of  $\alpha_{1}^{-1}$ compared with the previous case. If we employ the same configuration used in the discussion of the entanglement degradation, we cannot observe any entangled state being created, so $C[\rho(\tau)] = 0$ in this configuration.

\begin{figure}[H]
\centering
\includegraphics[scale=0.7]{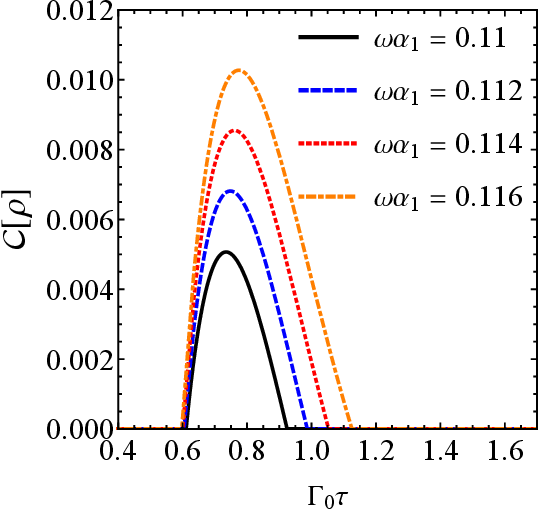}
\caption{Comparison of entanglement harvesting for different values of $\omega \alpha_1$. The initial state was prepared to be $\ket{E}$. We use  $\omega \alpha_2 = 1$.} 
\label{fig:ent_harv_e}
\end{figure}

When setting the initial state to be $\ket{E}$, we obtain the results illustrated in Fig. \ref{fig:ent_harv_e}. We observe that the entangled state is only created by significantly decreasing the acceleration of the second two-level system $\alpha_2^{-1}$. Moreover, the phenomenon of entanglement harvesting only occurs for small values of this acceleration. In this situation, we observe that we only have entangled states for $\alpha_2 \gg \alpha_1$ -- we observe that this entangled state has short lifetime. Furthermore, when we increase the value of $\alpha_{1}$, the associated entanglement increases. It is important to emphasize that although $\alpha_2$ is the parameter that most assists the entanglement harvesting, this phenomenon only occurs for relatively high values for $\alpha_1$ as well. This difficulty in creating entangled state is in agreement with the results discussed in Ref.~\cite{hu2015}.

\begin{figure}[H]
\centering
\includegraphics[scale=0.7]{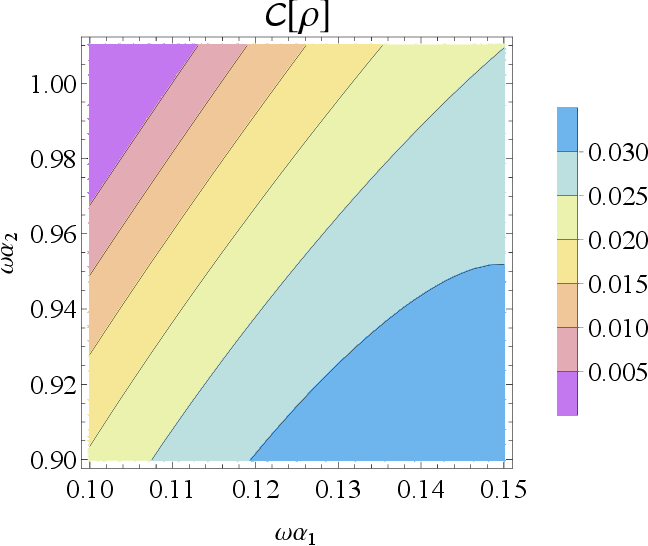}
\caption{Comparison of entanglement harvesting for different values of $\omega\alpha_1$ and $\omega\alpha_2$. The initial state was prepared to be $\ket{E}$ and we fixed the time parameter to $\Gamma_{0}\tau=0.75$.} 
\label{fig:ent_harv_accels}
\end{figure}

As claimed above, our generalized master equation  allows us to study how the entanglement of creation evolves when we scan through many different values for both proper accelerations. In Fig. \ref{fig:ent_harv_accels} we consider a fixed parameter $\Gamma_{0} \tau$ when the concurrence reaches its maximum value. To have a non-trivial result, we set values for $\alpha_2$ much greater than $\alpha_1$. An interesting work that studies how the acceleration can assist entanglement harvesting is presented in Ref.~\cite{liu2021does}. The authors investigate three different acceleration scenarios using the Unruh-DeWitt particle-detector model -- this is the main difference between the procedure that we have used here. This system configuration can now be revisited using the theory of quantum open systems and the master equation derived in this paper.


\section{Conclusions}\label{sec:conclusions}

Using the theory of open quantum system we were able to derive a generalized master equation for a system of two two-level system travelling along different world lines. We have also established a suitable generalized Lindblad-Kossakowski form and made use of the Kossakowski matrix formalism. We observe that this equation has a similar form of the one used for two-level systems in the same world lines. The crucial distinction here relies on the function given by Eq.~\eqref{eqq:gamma_gen}, which allows us to investigate how differences in the proper accelerations of the two-level systems affect entanglement dynamics.

We studied the phenomenon of entanglement degradation and observed that as the proper acceleration of the two-level systems decreases, the lifetime of the entangled state increases, in agreement with the Unruh-Davies effect. In particular, the antisymmetric entangled state presents a slower decoherence in comparison with the symmetric entangled state. In any case, we were able to observe the occurrence of entanglement sudden death in all situations under studied. On the other hand, we also observed that the entanglement harvesting is also possible in this scenario and we have investigated its dependence on the values of the proper acceleration of both two-level systems. We have demonstrated that a quantum system initially prepared in the ground state supports entanglement for a longer time in comparison with one initially prepared in the excited separable state. Furthermore, entanglement creation is favoured when the system undergoes small values of accelerations.

The formalism developed in this work provides us with a new way to understand entanglement dynamics in non-inertial reference frames.  The procedure used here can be used to study a more physical situation, for example a pair of dipoles interacting with the electromagnetic field, and even more complex situations, as with two-level systems in different spacetime regions in a Kerr black-hole background. These subjects are under investigation by the authors.

\section*{Acknowledgements}

We would like to thank C. A. D. Zarro and F. de Melo for useful discussions. This work was partially supported by Conselho Nacional de Desenvolvimento Cient\'{\i}fico e Tecnol\'{o}gico -- CNPq, under grants 303436/2015-8 (N.F.S.) and 317548/2021-2 (G.M.),~and Funda\c{c}\~ao Carlos Chagas Filho de Amparo \`a Pesquisa do Estado do Rio de Janeiro -- FAPERJ under grants E-26/202.725/2018 and E-26/201.142/2022 (G.M.).

\appendix
\section{Computation of the $\Gamma^{(\alpha \beta)}$ function}\label{app:gammas}
In order to calculate the function $\Gamma^{(\alpha \beta)}$ from Eq. \eqref{eqq:gamma_ii_explicit}, we have to compute the Fourier transform presented in Eq.~\eqref{eqq:gamma_ii_fourier}. Let us start with the $\alpha = \beta$ contribution. For simplicity we write $\Gamma^{(\beta \beta)}(a_{\beta} \omega) = \Gamma^{(\beta)}(\omega)$. We use the positive frequency Wightman function given by $\eqref{eqq:PWF_II}$ into Eq. \eqref{eqq:gamma_ii_fourier} and then we must solve the following integral
	\begin{align}\label{app:gamma_initial}
	 \Gamma^{(\beta)}( \omega) = - \frac{a_{\beta}^2}{16 \pi^2 \alpha_{\beta}^{2}}\int_{- \infty}^{\infty}ds~\frac{e^{i \omega a_{\beta} s}}{\sinh^2\left(\frac{1}{2 \alpha_{1}}(s)\right)}.
	\end{align}
The reason we omit the $2 i \epsilon$ term is that we will perform the following change of coordinates
	\begin{equation}
	s = r - i\pi \alpha_1.
	\end{equation}
Therefore, the integral in Eq. \eqref{app:gamma_initial} reads
	\begin{equation}\label{app:gamma_ii_inc}
	 \Gamma^{(\beta)}( \omega) = \frac{e^{\omega \alpha_{\beta} \pi}}{16 \pi^2 \alpha_{1}^{2}}\int_{-\infty}^{\infty}dr~\frac{\cos \left(\omega \alpha_{\beta} r\right) + i\sin \left(\omega \alpha_{\beta} r\right) }{\cosh^{2}\left(\frac{r}{2 \alpha_{1}}\right)}.
	\end{equation}
In Eq. \eqref{app:gamma_ii_inc}, the imaginary part is zero and the real contribution can be performed by making use of \cite{gradshteyn2014table}
	\begin{equation}
	\int_{0}^{\infty}\frac{\cos ax }{\cosh^{2}\beta x}dx = \frac{a\pi }{2\beta^2 \sinh \frac{a\pi}{2\beta}},
	\end{equation}
with $\realp \beta >0$ and $a>0$. Then, the $\Gamma^{(\beta)}(\omega)$ for $\omega>0$ can be written as
	\begin{equation}\label{app:gamma_ii_final}
	\Gamma^{(\beta)}(\omega) = \frac{\omega \alpha_{\beta}}{2 \pi \alpha_{1}}\left(\frac{1}{1 - e^{- 2\pi \omega \alpha_{\beta}}}\right).
	\end{equation}
For the case where $\omega<0$ we perform the change $\omega \rightarrow -\omega$	in Eq. \eqref{app:gamma_initial}. The following procedure is similar to the one used in the previous case. We observe that we obtain the same result as Eq. \eqref{app:gamma_ii_final} with $-\omega$ instead. Therefore, Eq. \eqref{app:gamma_ii_final} is valid for any value of $\omega$.

To evaluate the $\alpha \neq \beta$ case we will perform the Fourier transform presented in Eq. \eqref{eqq:gamma_ii_fourier}. By using Eq. \eqref{eqq:PWF_IJ} we have
	\begin{equation}\label{app:gamma_ij_initial}
	\Gamma^{(\alpha \beta)}( a_{\beta}\omega) = - \frac{a_{\alpha} a_{\beta}}{16\pi \alpha_1 \alpha_2}\int_{-\infty}^{\infty} ds \frac{e^{i \omega a_{\beta} s}}{\sinh\left(\frac{s}{2\alpha_{1}} + \frac{\phi}{2}\right)\sinh\left(\frac{s}{2\alpha_{1}} - \frac{\phi}{2}\right)},
	\end{equation}
where we again have omitted  the $i\epsilon$ term. We perform the following change of coordinates in Eq. \eqref{app:gamma_ij_initial}
	\begin{equation}
	s \rightarrow r - i \pi \alpha_{1}.
	\end{equation}
 Therefore we obtain
	\begin{equation}\label{app:gamma_ij_2}
	\Gamma^{(\alpha \beta)}( a_{\beta}\omega) = \frac{a_{\alpha} a_{\beta} e^{\pi \omega \alpha_{\beta}}}{8\pi^2 \alpha_1 \alpha_2} \int_{-\infty}^{\infty}dr \frac{\cos \left(\omega a_{\beta} r\right) + i\sin \left(\omega a_{\beta} r\right) }{\cos \left(\frac{r}{\alpha_1}\right) + \cosh \phi}.
	\end{equation}	 
Again, in Eq. \eqref{app:gamma_ij_2}, the imaginary part of this integral is zero and the real contribution can be performed by making use of \cite{gradshteyn2014table}
	\begin{equation}
	\int_{0}^{\infty} \frac{\cos ax }{b\cosh\beta x + c}dx = \frac{\pi \sin \left(\frac{a}{\beta} \text{arccosh} \frac{c}{b}\right) }{\beta \sqrt{c^2 - b^2} \sinh \frac{a\pi}{\beta}},
	\end{equation}
with $0<b<c$, $\realp{\beta}> 0$ and $a> 0$. Then, the $\Gamma^{(\alpha \beta)}(a_{\beta}\omega)$ for $\omega>0$ can be written as
	\begin{equation}\label{app:gamma_ij_semifinal}
	\Gamma^{(\alpha \beta)}(a_{\beta}\omega) = \frac{a_{\alpha}a_{\beta}}{2 \pi \alpha_{2} \sinh \phi}\left(\frac{\sin \left(\omega a_{\beta}\alpha_{1}\phi\right)}{1 - e^{- 2\pi \omega a_{\beta}\alpha_{1}}}\right).
	\end{equation}
As we have $\alpha \neq \beta$, $a_1 = 1$ and $a_2 = \alpha_{2}/\alpha_{1}$ then we can write $a_{\beta}\alpha_{1} = \alpha_{\beta}$ and Eq. \eqref{app:gamma_ij_semifinal} becomes
	\begin{equation}\label{app:gamma_ij_final}
	\Gamma^{(\alpha \beta)}(\omega) = \frac{\sin \left(\omega \alpha_{\beta}\phi\right)}{2 \pi \alpha_{1} \sinh \phi}\left(\frac{1}{1 - e^{- 2\pi \omega \alpha_{\beta}}}\right).
	\end{equation}
As can be observed, to obtain this function for $\omega < 0$ we simply perform the change $\omega \rightarrow -\omega$ in Eq. \eqref{app:gamma_ij_initial}. This change leads to the same result as Eq. \eqref{app:gamma_ij_final} with the same change in $\omega$. Therefore, Eq. \eqref{app:gamma_ij_final} is valid for any value of $\omega$.
\section{Explicit form of the generalized master equation}\label{app:gmeterms}

In this appendix we present the explicit form of the master equation \eqref{eqq:gme_scalar_field} where we consider the contribution only of the dissipative part of it as discussed in Sec. \ref{sec:ent_dynamics}. From Eq. \eqref{eqq:rho_initial_basis} we only have to compute eight components. These are given by the following set of differential equations:
\begin{align}
\frac{d\rho_{GG}(\tau)}{d\tau} =  &- 2 \left(A^{(11)} + A^{(22)} - B^{(11)} - B^{(22)}\right)\rho_{GG}(\tau)\nonumber\\
&+\left(A^{(11)} + A^{(22)} + B^{(11)} + B^{(22)}\right.\nonumber\\
 &- \left. A^{(12)} - A^{(21)} - B^{(12)} - B^{(21)}\right)\rho_{AA}(\tau)\nonumber\\
&+ \left(A^{(11)} - A^{(22)} + B^{(11)} - B^{(22)}\right.\nonumber\\
 &- \left. A^{(12)} + A^{(21)} + B^{(12)} - B^{(21)}\right)\rho_{AS}(\tau)\nonumber\\
 &+ \left(A^{(11)} - A^{(22)} + B^{(11)} - B^{(22)}\right.\nonumber\\
 &+ \left. A^{(12)} - A^{(21)} + B^{(12)} - B^{(21)}\right)\rho_{SA}(\tau)\nonumber\\
  &+ \left(A^{(11)} + A^{(22)} + B^{(11)} + B^{(22)}\right.\nonumber\\
 &+ \left. A^{(12)} + A^{(21)} + B^{(12)} + B^{(21)}\right)\rho_{SS}(\tau),\label{appeq-gme1}
\end{align}
\begin{align}  
 \frac{d\rho_{SS}(\tau)}{d\tau} =  &- 2 \left(A^{(11)} + A^{(22)} + A^{(12)} + A^{(21)}\right)\rho_{SS}(\tau)\nonumber\\
&+\left(  B^{(22)} - B^{(11)} + B^{(12)} - B^{(21)}\right)\rho_{AS}(\tau)\nonumber\\
&+ \left(A^{(11)} + A^{(22)} + B^{(11)} + B^{(22)}\right.\nonumber\\
 &+ \left. A^{(12)} + A^{(21)} + B^{(12)} + B^{(21)}\right)\rho_{EE}(\tau)\nonumber\\
 &+ \left(A^{(11)} + A^{(22)} - B^{(11)} - B^{(22)}\right.\nonumber\\
 &+ \left. A^{(12)} + A^{(21)} - B^{(12)} - B^{(21)}\right)\rho_{GG}(\tau)\nonumber\\
  &+ \left(  B^{(22)} - B^{(11)}  - B^{(12)} + B^{(21)}\right)\rho_{SA}(\tau),\label{appeq-gme2}
\end{align}
\begin{align}  
   \frac{d\rho_{AA}(\tau)}{d\tau} =  &- 2 \left(A^{(11)} + A^{(22)} - A^{(12)} - A^{(21)}\right)\rho_{AA}(\tau)\nonumber\\
&+\left(  B^{(22)} - B^{(11)} + B^{(12)} - B^{(21)}\right)\rho_{AS}(\tau)\nonumber\\
&+ \left(A^{(11)} + A^{(22)} + B^{(11)} + B^{(22)}\right.\nonumber\\
 &+ \left. A^{(12)} + A^{(21)} + B^{(12)} + B^{(21)}\right)\rho_{EE}(\tau)\nonumber\\
 &+ \left(A^{(11)} + A^{(22)} - B^{(11)} - B^{(22)}\right.\nonumber\\
 &+ \left. A^{(12)} + A^{(21)} - B^{(12)} - B^{(21)}\right)\rho_{GG}(\tau)\nonumber\\
  &+ \left(  B^{(22)} - B^{(11)}  - B^{(12)} + B^{(21)}\right)\rho_{SA}(\tau),\label{appeq-gme3}
 \end{align}
\begin{align}  
   \frac{d\rho_{EE}(\tau)}{d\tau} =  &- 2 \left(A^{(11)} + A^{(22)} + B^{(11)} + B^{(22)}\right)\rho_{EE}(\tau)\nonumber\\
&+ \left(A^{(22)} - A^{(11)} + B^{(11)} - B^{(22)}\right.\nonumber\\
 &- \left. A^{(12)} + A^{(21)} + B^{(12)} - B^{(21)}\right)\rho_{SA}(\tau)\nonumber\\
&+ \left(A^{(11)} + A^{(22)} - B^{(11)} - B^{(22)}\right.\nonumber\\
 &- \left. A^{(12)} - A^{(21)} + B^{(12)} + B^{(21)}\right)\rho_{AA}(\tau)\nonumber\\
 &+ \left(A^{(22)} - A^{(11)} + B^{(11)} - B^{(22)}\right.\nonumber\\
 &+ \left. A^{(12)} - A^{(21)} - B^{(12)} + B^{(21)}\right)\rho_{AS}(\tau)\nonumber\\
&+ \left(A^{(22)} + A^{(11)} - B^{(11)} - B^{(22)}\right.\nonumber\\
 &+ \left. A^{(12)} + A^{(21)} - B^{(12)} - B^{(21)}\right)\rho_{SS}(\tau),\label{appeq-gme4}  
\end{align}
\begin{align}  
   \frac{d\rho_{AS}(\tau)}{d\tau} =  &- 2 \left(A^{(11)} + A^{(22)}\right)\rho_{AS}(\tau)\nonumber\\
&+\left(  B^{(22)} - B^{(11)} - B^{(12)} + B^{(21)}\right)\rho_{AA}(\tau)\nonumber\\
&+ \left( A^{(22)}- A^{(11)}  - B^{(11)} + B^{(22)}\right.\nonumber\\
 &+ \left. A^{(12)} - A^{(21)} + B^{(12)} - B^{(21)}\right)\rho_{EE}(\tau)\nonumber\\
 &+ \left(A^{(11)} - A^{(22)} - B^{(11)} + B^{(22)}\right.\nonumber\\
 &- \left. A^{(12)} + A^{(21)} + B^{(12)} - B^{(21)}\right)\rho_{GG}(\tau)\nonumber\\
&+\left(  B^{(22)} - B^{(11)} - B^{(12)} + B^{(21)}\right)\rho_{SS}(\tau),\label{appeq-gme5}
\end{align}
\begin{align}  
   \frac{d\rho_{SA}(\tau)}{d\tau} =  &- 2 \left(A^{(11)} + A^{(22)}\right)\rho_{SA}(\tau)\nonumber\\
&+\left(  B^{(22)} - B^{(11)} + B^{(12)} - B^{(21)}\right)\rho_{AA}(\tau)\nonumber\\
&+ \left( A^{(22)}- A^{(11)}  - B^{(11)} + B^{(22)}\right.\nonumber\\
 &- \left. A^{(12)} + A^{(21)} - B^{(12)} + B^{(21)}\right)\rho_{EE}(\tau)\nonumber\\
 &+ \left(A^{(11)} - A^{(22)} - B^{(11)} + B^{(22)}\right.\nonumber\\
 &+ \left. A^{(12)} - A^{(21)} - B^{(12)} + B^{(21)}\right)\rho_{GG}(\tau)\nonumber\\
&+\left(  B^{(22)} - B^{(11)} + B^{(12)} - B^{(21)}\right)\rho_{SS}(\tau),\label{appeq-gme6}
\end{align}
\begin{align}
\frac{d\rho_{GE}(\tau)}{d\tau} =  &- 2 \left(A^{(11)} + A^{(22)}\right)\rho_{GE}(\tau)\label{appeq-gme7},\\
\frac{d\rho_{EG}(\tau)}{d\tau} =  &- 2 \left(A^{(11)} + A^{(22)}\right)\rho_{EG}(\tau).\label{appeq-gme8}
\end{align}

\bibliographystyle{unsrt}
\bibliography{references}

\begin{thebibliography}{100}

\bibitem{Bire82}
N.~D. Birell and P.~C.~W. Davies.
\newblock {\em Quantum Fields in Curved Space}.
\newblock Cambridge University Press, Cambridge, 1982.

\bibitem{Parker:2009uva}
Leonard~E. Parker and D.~Toms.
\newblock {\em {Quantum Field Theory in Curved Spacetime}}.
\newblock Cambridge University Press, 8 2009.

\bibitem{Davies:1974th}
P.~C.~W. Davies.
\newblock {Scalar particle production in Schwarzschild and Rindler metrics}.
\newblock {\em J. Phys. A}, 8:609--616, 1975.

\bibitem{Unruh:1976db}
W.~G. Unruh.
\newblock {Notes on black hole evaporation}.
\newblock {\em Phys. Rev. D}, 14:870, 1976.

\bibitem{unruh1984}
William~G Unruh and Robert~M Wald.
\newblock What happens when an accelerating observer detects a {R}indler
  particle.
\newblock {\em Phys. Rev. D}, 29(6):1047, 1984.

\bibitem{hawking75}
S.~W. Hawking.
\newblock Particle creation by black holes.
\newblock {\em Commun. Math. Phys.}, 43:199--220, Aug 1975.

\bibitem{yu08}
Hongwei Yu and Jialin Zhang.
\newblock Understanding hawking radiation in the framework of open quantum
  systems.
\newblock {\em Phys. Rev. D}, 77:024031, Jan 2008.

\bibitem{fulling1973}
Stephen~A Fulling.
\newblock Nonuniqueness of canonical field quantization in {R}iemannian
  space-time.
\newblock {\em Phys. Rev. D}, 7(10):2850, 1973.

\bibitem{Candelas:1976jv}
P.~Candelas and D.~J. Raine.
\newblock {Quantum Field Theory in Incomplete Manifolds}.
\newblock {\em J. Math. Phys.}, 17:2101--2112, 1976.

\bibitem{svaiter1992}
Bernar~F Svaiter and Nami~F Svaiter.
\newblock Inertial and noninertial particle detectors and vacuum fluctuations.
\newblock {\em Phys. Rev. D}, 46(12):5267, 1992.
\newblock [Erratum: Phys. Rev. D {\bf 47}, 4802(E) (1993)].

\bibitem{Ford:1994zz}
L.~H. Ford, N.~F. Svaiter, and Marcelo~L. Lyra.
\newblock {Radiative properties of a two-level system in the presence of
  mirrors}.
\newblock {\em Phys. Rev. A}, 49:1378--1386, 1994.

\bibitem{Shih-Yuin:06}
Shih-Yuin Lin and B.~L. Hu.
\newblock {Accelerated Detector - Quantum Field Correlations: From Vacuum
  Fluctuations to Radiation Flux}.
\newblock {\em Phys. Rev. D}, 73:124018, 2006.

\bibitem{Crispino:2007eb}
L.~C.~B. Crispino, A.~Higuchi, and G.~E.~A. Matsas.
\newblock {The Unruh effect and its applications}.
\newblock {\em Rev. Mod. Phys.}, 80:787, 2008.

\bibitem{Lenz:2010vn}
F.~Lenz, K.~Ohta, and K.~Yazaki.
\newblock {Static interactions and stability of matter in Rindler space}.
\newblock {\em Phys. Rev. D}, 83:064037, 2011.

\bibitem{Arias:2011yg}
E.~Arias, G.~Krein, G.~Menezes, and N.~F. Svaiter.
\newblock {Thermal Radiation from a Fluctuating Event Horizon}.
\newblock {\em Int. J. Mod. Phys. A}, 27:1250129, 2012.

\bibitem{Menezes:2015iva}
G.~Menezes and N.~F. Svaiter.
\newblock {Radiative processes of uniformly accelerated entangled atoms}.
\newblock {\em Phys. Rev. A}, 93(5):052117, 2016.

\bibitem{Ben-Benjamin:2019opz}
J.~S. Ben-Benjamin et~al.
\newblock {Unruh Acceleration Radiation Revisited}.
\newblock {\em Int. J. Mod. Phys. A}, 34(28):1941005, 2019.

\bibitem{Rodriguez-Camargo:2016fbq}
C.~D. Rodr\'\i{}guez-Camargo, N.~F. Svaiter, and G.~Menezes.
\newblock Finite-time response function of uniformly accelerated entangled
  atoms.
\newblock {\em Annals Phys.}, 396:266--291, 2018.

\bibitem{Arias:2015moa}
E.~Arias, J.~G. Due\~nas, G.~Menezes, and N.~F. Svaiter.
\newblock Boundary effects on radiative processes of two entangled atoms.
\newblock {\em JHEP}, 07:147, 2016.

\bibitem{Fewster:2016ewy}
Christopher~J. Fewster, Benito~A. Ju\'arez-Aubry, and Jorma Louko.
\newblock {Waiting for Unruh}.
\newblock {\em Class. Quant. Grav.}, 33(16):165003, 2016.

\bibitem{Menezes:2016quu}
G.~Menezes.
\newblock {Spontaneous excitation of an atom in a Kerr spacetime}.
\newblock {\em Phys. Rev. D}, 95(6):065015, 2017.
\newblock [Erratum: Phys.Rev.D 97, 029901(E) (2018)].

\bibitem{Menezes:2017rby}
G.~Menezes, N.~F. Svaiter, and C.~A.~D. Zarro.
\newblock Entanglement dynamics in random media.
\newblock {\em Phys. Rev. A}, 96(6):062119, 2017.

\bibitem{Picanco:2020api}
Gabriel Pican\c{c}o, Nami~F. Svaiter, and Carlos A.~D. Zarro.
\newblock Radiative processes of entangled detectors in rotating frames.
\newblock {\em JHEP}, 08:025, 2020.

\bibitem{PhysRevD.101.025009}
Wenting Zhou and Hongwei Yu.
\newblock Radiation-reaction-induced transitions of two maximally entangled
  atoms in noninertial motion.
\newblock {\em Phys. Rev. D}, 101:025009, Jan 2020.

\bibitem{Tjoa:2022oxv}
Erickson Tjoa and Robert~B. Mann.
\newblock {Unruh-DeWitt detector in dimensionally-reduced static spherically
  symmetric spacetimes}.
\newblock {\em JHEP}, 03:014, 2022.

\bibitem{Hu:2022nxc}
Hui Hu, Jialin Zhang, and Hongwei Yu.
\newblock {Harvesting Entanglement by non-identical detectors with different
  energy gaps}.
\newblock 4 2022.

\bibitem{kaplanek2020hot}
Greg Kaplanek and CP~Burgess.
\newblock Hot accelerated qubits: decoherence, thermalization, secular growth
  and reliable late-time predictions.
\newblock {\em Journal of High Energy Physics}, 2020(3):1--49, 2020.

\bibitem{arrechea2021inversion}
Julio Arrechea, Carlos Barcel{\'o}, Luis~J Garay, and Gerardo
  Garc{\'\i}a-Moreno.
\newblock Inversion of statistics and thermalization in the unruh effect.
\newblock {\em Physical Review D}, 104(6):065004, 2021.

\bibitem{carballo2019unruh}
Ra{\'u}l Carballo-Rubio, Luis~J Garay, Eduardo Mart{\'\i}n-Mart{\'\i}nez, and
  Jos{\'e} De~Ram{\'o}n.
\newblock Unruh effect without thermality.
\newblock {\em Physical review letters}, 123(4):041601, 2019.

\bibitem{juarez2014onset}
Benito~A Ju{\'a}rez-Aubry and Jorma Louko.
\newblock Onset and decay of the 1+ 1 hawking--unruh effect: what the
  derivative-coupling detector saw.
\newblock {\em Classical and Quantum Gravity}, 31(24):245007, 2014.

\bibitem{PhysRevD.95.025020}
Dimitris Moustos and Charis Anastopoulos.
\newblock Non-markovian time evolution of an accelerated qubit.
\newblock {\em Phys. Rev. D}, 95:025020, Jan 2017.

\bibitem{dewitt1975}
Bryce~S DeWitt.
\newblock Quantum {F}ield {T}heory in {C}urved {S}pacetime.
\newblock {\em Phys. Rep.}, 19(6):295--357, 1975.

\bibitem{Hawking:1979ig}
S.~W. Hawking and W.~Israel.
\newblock {\em {General Relativity}: {An Einstein Centenary Survey}}.
\newblock Univ. Pr., Cambridge, UK, 1979.

\bibitem{glauber1963A}
Roy~J Glauber.
\newblock The quantum theory of optical coherence.
\newblock {\em Phys. Rev.}, 130(6):2529, 1963.

\bibitem{soares21}
M.~S. Soares, N.~F. Svaiter, C.~A.~D. Zarro, and G.~Menezes.
\newblock Uniformly accelerated quantum counting detector in minkowski and
  fulling vacuum states.
\newblock {\em Phys. Rev. A}, 103:042225, Apr 2021.

\bibitem{Martin-Martinez:2014qda}
Eduardo Martin-Martinez and Jorma Louko.
\newblock {Particle detectors and the zero mode of a quantum field}.
\newblock {\em Phys. Rev. D}, 90(2):024015, 2014.

\bibitem{Martin-Martinez:2015psa}
Eduardo Martin-Martinez.
\newblock {Causality issues of particle detector models in QFT and Quantum
  Optics}.
\newblock {\em Phys. Rev. D}, 92(10):104019, 2015.

\bibitem{Hummer:2015xaa}
Daniel H\"ummer, Eduardo Martin-Martinez, and Achim Kempf.
\newblock {Renormalized Unruh-DeWitt Particle Detector Models for Boson and
  Fermion Fields}.
\newblock {\em Phys. Rev. D}, 93(2):024019, 2016.

\bibitem{Tjoa:2021bcn}
Erickson Tjoa, Irene Lopez-Guti\'errez, Allison Sachs, and Eduardo
  Mart\'\i{}n-Mart\'\i{}nez.
\newblock {What makes a particle detector click}.
\newblock {\em Phys. Rev. D}, 103(12):125021, 2021.

\bibitem{deRamon:2021nry}
Jos\'e de~Ram\'on, Maria Papageorgiou, and Eduardo Mart\'\i{}n-Mart\'\i{}nez.
\newblock {Relativistic causality in particle detector models:
  Faster-than-light signaling and impossible measurements}.
\newblock {\em Phys. Rev. D}, 103(8):085002, 2021.

\bibitem{nielsen2011}
Michael~A Nielsen and Isaac Chuang.
\newblock {\em Quantum computation and quantum information}.
\newblock Cambridge University Press, 2010.

\bibitem{Plenio:1998wq}
M.~B. Plenio, S.~F. Huelga, A.~Beige, and P.~L. Knight.
\newblock {Cavity loss induced generation of entangled atoms}.
\newblock {\em Phys. Rev. A}, 59:2468--2475, 1999.

\bibitem{cite-key}
A.~M. Basharov.
\newblock Entanglement of atomic states upon collective radiative decay.
\newblock {\em JETP Letters}, 75(3):123--126, 2002.

\bibitem{doi:10.1080/09500340308234584}
Z.~Ficek and R.~Tanas.
\newblock Entanglement induced by spontaneous emission in spatially extended
  two-atom systems.
\newblock {\em J. Mod. Opt.}, 50(18):2765--2779, 2003.

\bibitem{Tana_2004}
R~Tanas and Z~Ficek.
\newblock Entangling two atoms via spontaneous emission.
\newblock {\em J. opt., B Quantum semiclass. opt.}, 6(3):S90--S97, mar 2004.

\bibitem{Amico:2007ag}
Luigi Amico, Rosario Fazio, Andreas Osterloh, and Vlatko Vedral.
\newblock {Entanglement in many-body systems}.
\newblock {\em Rev. Mod. Phys.}, 80:517--576, 2008.

\bibitem{Peres:2002ip}
Asher Peres, Petra~F. Scudo, and Daniel~R. Terno.
\newblock {Quantum entropy and special relativity}.
\newblock {\em Phys. Rev. Lett.}, 88:230402, 2002.

\bibitem{Gingrich:2002ota}
Robert~M. Gingrich and Christoph Adami.
\newblock {Quantum Entanglement of Moving Bodies}.
\newblock {\em Phys. Rev. Lett.}, 89(27):270402, 2002.

\bibitem{Gingrich:2002otaa}
Robert~M. Gingrich, Attila~J. Bergou, and Christoph Adami.
\newblock {Entangled light in moving frames}.
\newblock {\em Phys. Rev. A}, 68:042102, 2003.

\bibitem{Shi:2004yt}
Yu~Shi.
\newblock {Entanglement in relativistic quantum field theory}.
\newblock {\em Phys. Rev. D}, 70:105001, 2004.

\bibitem{Czachor}
Marek Czachor.
\newblock {Comment on ``Quantum Entropy and Special Relativity"}.
\newblock {\em Phys. Rev. Lett.}, 94:078901, 2005.

\bibitem{Jordan:2006kt}
Thomas~F. Jordan, Anil Shaji, and E.~C.~G. Sudarshan.
\newblock {Lorentz transformations that entangle spins and entangle momenta}.
\newblock {\em Phys. Rev. A}, 75:022101, 2007.

\bibitem{Adesso:2007wi}
Gerardo Adesso, Ivette Fuentes-Schuller, and Marie Ericsson.
\newblock {Continuous variable entanglement sharing in non-inertial frames}.
\newblock {\em Phys. Rev. A}, 76:062112, 2007.

\bibitem{Datta}
Animesh Datta.
\newblock {Quantum discord between relatively accelerated observers}.
\newblock {\em Phys. Rev. A}, 80:052304, 2009.

\bibitem{Peres:2002wx}
Asher Peres and Daniel~R. Terno.
\newblock {Quantum information and relativity theory}.
\newblock {\em Rev. Mod. Phys.}, 76:93--123, 2004.

\bibitem{Lin:2010zzb}
Shih-Yuin Lin and B.~L. Hu.
\newblock {Entanglement creation between two causally disconnected objects}.
\newblock {\em Phys. Rev. D}, 81:045019, 2010.

\bibitem{Bruschi:2010mc}
David~E. Bruschi, Jorma Louko, Eduardo Martin-Martinez, Andrzej Dragan, and
  Ivette Fuentes.
\newblock {The Unruh effect in quantum information beyond the single-mode
  approximation}.
\newblock {\em Phys. Rev. A}, 82:042332, 2010.

\bibitem{Ostapchuk:2011ud}
David C.~M. Ostapchuk, Shih-Yuin Lin, Robert~B. Mann, and B.~L. Hu.
\newblock {Entanglement Dynamics between Inertial and Non-uniformly Accelerated
  Detectors}.
\newblock {\em JHEP}, 07:072, 2012.

\bibitem{Hu:2012jr}
B.~L. Hu, Shih-Yuin Lin, and Jorma Louko.
\newblock {Relativistic Quantum Information in Detectors-Field Interactions}.
\newblock {\em Class. Quant. Grav.}, 29:224005, 2012.

\bibitem{CQG12}
{Relativistic quantum information, special issue of Classical and Quantum
  Gravity}.
\newblock {\em Class. Quantum Grav.}, 29:(11), 2012.

\bibitem{Lin:2015aua}
Shih-Yuin Lin, Chung-Hsien Chou, and Bei-Lok Hu.
\newblock {Entanglement Dynamics of Detectors in an Einstein Cylinder}.
\newblock {\em JHEP}, 03:047, 2016.

\bibitem{Martin-Martinez:2010yva}
Eduardo Martin-Martinez, Luis~J. Garay, and Juan Leon.
\newblock {Unveiling quantum entanglement degradation near a Schwarzschild
  black hole}.
\newblock {\em Phys. Rev. D}, 82:064006, 2010.

\bibitem{yu2016pra}
Yiquan Yang, Jiawei Hu, and Hongwei Yu.
\newblock Entanglement dynamics for uniformly accelerated two-level atoms
  coupled with electromagnetic vacuum fluctuations.
\newblock {\em Phys. Rev. A}, 94:032337, Sep 2016.

\bibitem{She:2019hjv}
Jiaozhen She, Jiawei Hu, and Hongwei Yu.
\newblock {Entanglement dynamics for circularly accelerated two-level atoms
  coupled with electromagnetic vacuum fluctuations}.
\newblock {\em Phys. Rev. D}, 99(10):105009, 2019.

\bibitem{He:2020xhz}
Pingyang He, Hongwei Yu, and Jiawei Hu.
\newblock {Entanglement dynamics for static two-level atoms in cosmic string
  spacetime}.
\newblock {\em Eur. Phys. J. C}, 80(2):134, 2020.

\bibitem{Bhattacharya:2021zgd}
Sourav Bhattacharya and Nitin Joshi.
\newblock {Entanglement degradation in multi-event horizon spacetimes}.
\newblock {\em Phys. Rev. D}, 105(6):065007, 2022.

\bibitem{martinmartinez-PhysRevD.92.064042}
Alejandro Pozas-Kerstjens and Eduardo Mart\'{\i}n-Mart\'{\i}nez.
\newblock Harvesting correlations from the quantum vacuum.
\newblock {\em Phys. Rev. D}, 92:064042, Sep 2015.

\bibitem{Sachs:2017exo}
Allison Sachs, Robert~B. Mann, and Eduardo Martin-Martinez.
\newblock {Entanglement harvesting and divergences in quadratic Unruh-DeWitt
  detector pairs}.
\newblock {\em Phys. Rev. D}, 96(8):085012, 2017.

\bibitem{Pozas-Kerstjens:2017xjr}
Alejandro Pozas-Kerstjens, Jorma Louko, and Eduardo Martin-Martinez.
\newblock {Degenerate detectors are unable to harvest spacelike entanglement}.
\newblock {\em Phys. Rev. D}, 95(10):105009, 2017.

\bibitem{MM-PhysRevD.94.064074}
Alejandro Pozas-Kerstjens and Eduardo Mart\'{\i}n-Mart\'{\i}nez.
\newblock Entanglement harvesting from the electromagnetic vacuum with
  hydrogenlike atoms.
\newblock {\em Phys. Rev. D}, 94:064074, Sep 2016.

\bibitem{Liu:2020jaj}
Zhihong Liu, Jialin Zhang, and Hongwei Yu.
\newblock {Entanglement harvesting in the presence of a reflecting boundary}.
\newblock {\em JHEP}, 08:020, 2021.

\bibitem{Tjoa:2020riy}
Erickson Tjoa and Eduardo Mart\'\i{}n-Mart\'\i{}nez.
\newblock {Vacuum entanglement harvesting with a zero mode}.
\newblock {\em Phys. Rev. D}, 101(12):125020, 2020.

\bibitem{Perche:2021clp}
T.~Rick Perche, Caroline Lima, and Eduardo Mart\'\i{}n-Mart\'\i{}nez.
\newblock {Harvesting entanglement from complex scalar and fermionic fields
  with linearly coupled particle detectors}.
\newblock {\em Phys. Rev. D}, 105(6):065016, 2022.

\bibitem{martinmartinez_prd21}
Erickson Tjoa and Eduardo Mart\'{\i}n-Mart\'{\i}nez.
\newblock When entanglement harvesting is not really harvesting.
\newblock {\em Phys. Rev. D}, 104:125005, Dec 2021.

\bibitem{Fuentes-Schuller:2004iaz}
Ivette Fuentes-Schuller and Robert~B. Mann.
\newblock {Alice falls into a black hole: Entanglement in non-inertial frames}.
\newblock {\em Phys. Rev. Lett.}, 95:120404, 2005.

\bibitem{mann2008quantum}
RB~Mann and I~Fuentes-Schuller.
\newblock Quantum entanglement in noninertial frames.
\newblock {\em Physics Essays}, 21(1):1, 2008.

\bibitem{Alsing_2012}
Paul~M Alsing and Ivette Fuentes.
\newblock Observer-dependent entanglement.
\newblock {\em Class. Quantum Grav.}, 29(22):224001, oct 2012.

\bibitem{Salton_2015}
Grant Salton, Robert~B Mann, and Nicolas~C Menicucci.
\newblock Acceleration-assisted entanglement harvesting and rangefinding.
\newblock {\em New J. Phys.}, 17(3):035001, mar 2015.

\bibitem{PhysRevD.104.025001}
Kensuke Gallock-Yoshimura, Erickson Tjoa, and Robert~B. Mann.
\newblock Harvesting entanglement with detectors freely falling into a black
  hole.
\newblock {\em Phys. Rev. D}, 104:025001, Jul 2021.

\bibitem{PhysRevD.92.064042}
Alejandro Pozas-Kerstjens and Eduardo Mart\'{\i}n-Mart\'{\i}nez.
\newblock Harvesting correlations from the quantum vacuum.
\newblock {\em Phys. Rev. D}, 92:064042, Sep 2015.

\bibitem{PhysRevA.100.062126}
Matheus~H. Zambianco, Andr\'e G.~S. Landulfo, and George E.~A. Matsas.
\newblock Observer dependence of entanglement in nonrelativistic quantum
  mechanics.
\newblock {\em Phys. Rev. A}, 100:062126, Dec 2019.

\bibitem{PhysRevD.102.065013}
Jialin Zhang and Hongwei Yu.
\newblock Entanglement harvesting for unruh-dewitt detectors in circular
  motion.
\newblock {\em Phys. Rev. D}, 102:065013, Sep 2020.

\bibitem{cong2020effects}
Wan Cong, Chen Qian, Michael~RR Good, and Robert~B Mann.
\newblock Effects of horizons on entanglement harvesting.
\newblock {\em JHEP}, 2020(10):1--19, 2020.

\bibitem{liu2021does}
Zhihong Liu, Jialin Zhang, Robert~B Mann, and Hongwei Yu.
\newblock Does acceleration assist entanglement harvesting?
\newblock {\em arXiv preprint arXiv:2111.04392}, 2021.

\bibitem{Menezes:2015uaa}
G.~Menezes and N.~F. Svaiter.
\newblock {Vacuum fluctuations and radiation reaction in radiative processes of
  entangled states}.
\newblock {\em Phys. Rev. A}, 92(6):062131, 2015.

\bibitem{Menezes:2015veo}
G.~Menezes.
\newblock {Radiative processes of two entangled atoms outside a Schwarzschild
  black hole}.
\newblock {\em Phys. Rev. D}, 94(10):105008, 2016.

\bibitem{Menezes:2017oeb}
G.~Menezes.
\newblock {Entanglement dynamics in a Kerr spacetime}.
\newblock {\em Phys. Rev. D}, 97(8):085021, 2018.

\bibitem{benatti04_pra}
F.~Benatti and R.~Floreanini.
\newblock Entanglement generation in uniformly accelerating atoms:
  Reexamination of the unruh effect.
\newblock {\em Phys. Rev. A}, 70:012112, Jul 2004.

\bibitem{benatti2010}
F.~Benatti, R.~Floreanini, and U.~Marzolino.
\newblock Entangling two unequal atoms through a common bath.
\newblock {\em Phys. Rev. A}, 81:012105, Jan 2010.

\bibitem{louko_prd_2020}
Boris Sokolov, Jorma Louko, Sabrina Maniscalco, and Iiro Vilja.
\newblock Unruh effect and information flow.
\newblock {\em Phys. Rev. D}, 101:024047, Jan 2020.

\bibitem{hu2015}
Jiawei Hu and Hongwei Yu.
\newblock Entanglement dynamics for uniformly accelerated two-level atoms.
\newblock {\em Phys. Rev. A}, 91:012327, Jan 2015.

\bibitem{breuer2002theory}
Heinz-Peter Breuer, Francesco Petruccione, et~al.
\newblock {\em The theory of open quantum systems}.
\newblock Oxford University Press on Demand, 2002.

\bibitem{tutorial}
Stanislaw Kryszewski and Justyna Czechowska-Kryszk.
\newblock {Master equation - tutorial approach}.
\newblock 2008.

\bibitem{cohen1986quantum}
Claude Cohen-Tannoudji, Bernard Diu, and Frank Laloe.
\newblock Quantum mechanics, volume 2.
\newblock {\em Quantum Mechanics}, 2:626, 1986.

\bibitem{wootters98}
William~K. Wootters.
\newblock Entanglement of formation of an arbitrary state of two qubits.
\newblock {\em Phys. Rev. Lett.}, 80:2245--2248, Mar 1998.

\bibitem{lindblad1976generators}
Goran Lindblad.
\newblock On the generators of quantum dynamical semigroups.
\newblock {\em Communications in Mathematical Physics}, 48(2):119--130, 1976.

\bibitem{gorini1976completely}
Vittorio Gorini, Andrzej Kossakowski, and Ennackal Chandy~George Sudarshan.
\newblock Completely positive dynamical semigroups of n-level systems.
\newblock {\em Journal of Mathematical Physics}, 17(5):821--825, 1976.

\bibitem{FICEK2002369}
Zbigniew Ficek and Ryszard Tanas.
\newblock Entangled states and collective nonclassical effects in two-atom
  systems.
\newblock {\em Phys. Rep.}, 372(5):369--443, 2002.

\bibitem{gallock2021entangled}
Kensuke Gallock-Yoshimura and Robert~B. Mann.
\newblock Entangled detectors nonperturbatively harvest mutual information.
\newblock {\em Phys. Rev. D}, 104:125017, Dec 2021.

\bibitem{doi:10.1126/science.1167343}
Ting Yu and J.~H. Eberly.
\newblock Sudden death of entanglement.
\newblock {\em Science}, 323(5914):598--601, 2009.

\bibitem{gradshteyn2014table}
Izrail~Solomonovich Gradshteyn and Iosif~Moiseevich Ryzhik.
\newblock {\em Table of integrals, series, and products}.
\newblock Academic press, 2014.

\end{thebibliography}
\end{document}